\documentclass{aastex631}

\newcounter{daggerfootnote}
\newcommand*{\daggerfootnote}[1]{%
   \setcounter{daggerfootnote}{\value{footnote}}%
    \renewcommand*{\thefootnote}{\fnsymbol{footnote}}%
    \footnote[2]{#1}%
    \setcounter{footnote}{\value{daggerfootnote}}%
    \renewcommand*{\thefootnote}{\arabic{footnote}}%
    }

\begin{document}

\def\RSUN{R$_{\sun}$}
\def\kms{$\rm km~s^{-1}$}
\def\cmcube{$\rm cm^{-3$}$}
\def\RSUNs{R$_{\sun}$ }
\def\kmss{$\rm km~s^{-1}$ }
\def\cmcubes{$\rm cm^{-3$}$ }
\def\Qbar{$<\rm Q>$}
\def\Qbars{$<\rm Q>$ }

\title{Heating of a plasma sheet in nonequilibrium ionization with nonthermal electrons}

\author[0000-0001-6412-5556]{Jin-Yi Lee}
\affiliation{Department of Astronomy and Space Science, Kyung Hee University, Yongin-si, Gyeonggi-do, 17104, Republic of Korea}
\email{jlee@khu.ac.kr}

\author[0000-0002-7868-1622]{John C. Raymond}
\affiliation{Center for Astrophysics $\vert$ Harvard $\&$ Smithsonian, 
Cambridge, MA 02138, USA}

\author[0000-0002-6903-6832]{Katharine K. Reeves}
\affiliation{Center for Astrophysics $\vert$ Harvard $\&$ Smithsonian, 
Cambridge, MA 02138, USA}

\author[0000-0002-9258-4490]{Chengcai Shen}
\affiliation{Yunnan Observatories, Chinese Academy of Sciences, P.O. Box 110, Kunming, Yunnan 650216, China}
\affiliation{Yunnan Key Laboratory of Solar Physics and Space Science, Kunming, Yunnan 650216, China}

\author{Stephen Kahler$^\dagger$}
\daggerfootnote{Passed away February 5, 2023}
\affiliation{Air Force Research Laboratory, Space Vehicles Directorate, 3550 Aberdeen Ave., Kirtland AFB, NM 87117, USA}

\author[0000-0001-6216-6944]{Yong-Jae Moon}
\affiliation{Department of Astronomy and Space Science, Kyung Hee University, Yongin-si, Gyeonggi-do, 17104, Republic of Korea}
\affiliation{G-LAMP NEXUS Institute/School of Space Research, Kyung Hee University,  Yongin, 17104, Republic of Korea}

\author[0000-0001-5900-6237]{Yeon-Han Kim}
\affiliation{Korea Astronomy \& Space Science Institute, 
Daejeon, 34055, Republic of Korea}



\begin{abstract}

A flux rope eruption on September 10, 2017 provides unique observations of the plasma sheet beneath the rising flux rope. 
The plasma sheet is likely in a nonequilibrium state in terms of both ionization and the electron distribution function. 
We trace the evolution of a blob in the plasma sheet using observations from the Atmospheric Imaging Assembly onboard the Solar Dynamics Observatory. 
We investigate the heating of plasma sheet material in the presence of non-Maxwellian electron distributions and nonequilibrium ionization. 
Our models compute time-dependent ion fractions, incorporating impulsive heating to various peak temperatures, continuous heating rates, and $\kappa$ values that represent the non-Maxwellian distribution.
The statistically preferred models constrain the effective impulsive heating temperature to above 20~MK. 
High-temperature solutions are permitted only for very low $\kappa$ values, indicating that suprathermal electrons play a significant role. 
Impulsive heating dominates the energy budget, with continuous heating contributing approximately 6\%–50\% of the initial impulsive energy input.

\end{abstract}


\section{Introduction} \label{sec:intro}

Magnetic reconnection is a fundamental plasma process that rapidly converts magnetic energy into 
kinetic energy, heating, and particle acceleration throughout astrophysical and space plasmas. 
Reconnection plays a key role in solar flares, coronal mass ejections, magnetospheric substorms, 
and other energetic events in astrophysics, where sudden magnetic topological changes release the energy \citep[e.g.,][]{priest02, Hesse2020, palmroth23}. Over the past decades, solar eruptions have been extensively simulated using magnetohydrodynamic (MHD) models (\citealp{kopp1976, lin00, antiochos1999, titov1999}; see also references in \citealp{priest02}), which show that the eruption forms a current sheet where magnetic reconnection takes place.

Energy conversion through magnetic reconnection in current sheets has been investigated using MHD models \citep{mei12, reeves10, reeves19, ye19,ye23, shen11, shen22, shen23}. Both \citet{reeves19} and \citet{ye23} find that the adiabatic and thermal conduction terms are important contributors to heating the current sheet plasma.  \citet{ye23} take into account turbulence and plasmoids generated during eruptions within the current sheets and find that the plasmoid reconnection dominates in the lower current sheet while turbulent reconnection dominates in the upper current sheet. As a result of magnetic reconnection, the plasma in current sheets is expected to be in nonequilibrium states, exhibiting both nonequilibrium ionization (NEI) and nonthermal electron velocity distributions due to rapid heating and particle accelerations \citep{bradshaw13, dudik17}. 

An earlier study shows that a model of Petschek-like reconnection in a post-coronal mass ejection current sheet predicts ionic fractions far from ionization equilibrium. The model also successfully reproduces the UV and X-ray emissions of a plasma sheet observed by the ultraviolet coronagraph spectrometer (UVCS) on board {\it{Solar and Heliospheric Observatory}} and by the X-ray telescope (XRT) on board {\it{Hinode}} \citep{ko10}. Another study on the NEI effects of Fe transition lines on magnetic reconnection in the solar corona shows that these effects are significant and sensitive to the electron densities below 10$^{10}$cm$^{-3}$ \citep{imada11}. NEI modeling combined with MHD simulations has also emphasized that NEI strongly affects the observed band ratios from the Atmospheric Imaging Assembly (AIA) on board {\it{Solar Dynamic Observatory}} \citep{shen13, shen23} and from the EUV Imaging Spectrometer (EIS) on board {\it{Hinode}} \citep{wraback25}. 

Nonthermal electrons accelerated in solar eruptions, such as solar flares, have been identified through hard X-ray observations \citep[e.g.,][]{lin1976, lin1981, oka15, Battaglia19, massa22}. The observed energy spectra show high-energy tails that indicate the presence of accelerated nonthermal electron populations \citep[see reviews in][]{oka18}. \citet{li22} find that the higher-energy component of the power-law spectrum results from the particles trapped in the current sheet, based on their test-particle study of particles in an evolving
reconnecting current sheet. \citet{chen20a} present a nonthermal microwave source at the location of the
central current sheet that extends upward to encompass a flux rope cavity. An MHD simulation by \citet{lix25} shows that particles accelerated by flare reconnection can be released through interchange reconnection between closed and open field lines, and these particles may become an important seed population for further acceleration by CME-driven shocks. In agreement with this view, \citet{kahler19} show that suprathermal particles measured in situ at 1 AU are well correlated with the seed particles accelerated near the Sun to E \textgreater 10 MeV. This suggests that particles accelerated in current sheets may contribute to the suprathermal populations that might be later reaccelerated in large SEP events. 

Direct observation of plasma sheets (sometimes referred to current sheets in observations, though current can not be measured in coronal images) is challenging due to their small thickness and low density structure. Several such events have been observed, mostly by UVCS and also by EIS, XRT, and AIA \citep{ciaravella02, ko03, bemporad06, lee06, ko10, savage10, landi12, liu13}. Towards the end of the last decade, an exceptional flux rope eruption event suitable for current sheet studies was observed on 2017 September 10, in which the eruption clearly revealed the current sheet structure underneath the erupting flux rope on the solar west limb. 
The event has been investigated through various approaches, studies on morphology and propagation kinematics \citep{yan18, long18, seaton18, gopalswamy18, veronig18}, plasma properties using UV and EUV observations \citep{doschek18, warren18, longcope18, polito18, li18, cai19, cai22, reeves20, imada21, gou24, Kittrell25}, nonthermal properties using microwave and hard X-ray observations \citep{gary18, chen20a, chen20b, chen21, li22}, and turbulence and blob observations within the plasma sheet \citep{cheng18, lee20, patel20, Xie2024}. This event was extremely energetic, producing a ground level enhancement (GLE) event at Earth \citep{omodei18, augusto19, liu23} and, it was detected even at Mars \citep{elrod18, lee18, guo18}.

In this work, we calculate the ion fractions in nonequilibrium states that include both NEI and nonthermal electrons for a current sheet observation on 2017 September 10. \citet{imada21} showed that the observed Fe~XXIV/Fe~XXIII ratios are consistent with NEI calculations assuming a constant electron temperature of 25~MK and a density of 10$^{10}$cm$^{-3}$ along the downstream region of the reconnection outflow in the current sheet observations. He also found that the nonthermal velocity increases significantly with height in the current sheet, which is largely consistent with the results of \citet{warren18}. A Parker transport model \citep{lix22} combined with an MHD simulation for this event shows that electrons are accelerated up to several MeV and fill a large volume of the reconnection region, as shown by microwave observations \citep{chen20b}. We model the ion fractions using a time-dependent ionization model \citep{shen15}, applying a $\kappa$ distribution \citep{cui19} representing high-energy tails deviating from a Maxwellian velocity distribution. Then, we compare the synthesized images in EUV with the AIA observations. 

\citet{warren18} measured plasma parameters assuming ionization equilibrium
and Maxwellian electron distributions. \citet{imada21} measured plasma parameters 
as a function of height allowing time-dependent ionization. \citet{polito18} 
measured line profiles that implied $\kappa < $3 if they were interpreted as
nonthermal velocity distributions. We confirm the basic results of \citet{warren18} 
and \citet{imada21} that the temperature is around 20 MK if the electrons are Maxwellian,
and the result of \citet{polito18} that  significantly higher temperatures are allowed if
the electrons are strongly non-Maxwellian. We add the new result that the
energy added as the blob moves up the plasma sheet is a modest fraction of
the energy added impulsively when the blob is formed. 

In Section 2, we describe plasma tracking and the calculation of the density along the observed plasma sheet, which is used as input to the nonequilibrium model. In Section 3, we introduce the heating model for the temperature evolution and the computation of ion fractions with nonthermal electrons, as well as the synthesis of AIA observations and the model selection using covariance-weighted $\chi^2$. In Sections 4 and 5, we present the results and discuss the relevant physical processes. In Section 6, we summarizes our conclusions.

\section{Data} \label{sec:data} 

Using a uniquely well-observed plasma sheet located beneath the rising flux rope, we investigate the plasma heating 
that is thought to occur in the magnetic reconnection region. 
In this section, we describe how we determine the time-dependent density of the plasma sheet, as well as 
the temperature and density of the background emission. 
For the plasma sheet temperature, we construct a model with impulsive and continuous heating, as described in Section 3. 
These quantities are required to examine the ion fractions under the NEI condition and to characterize the non-thermal electron distribution. 

\subsection{2017 September 10 event} \label{subsec:event} 

A flux rope eruption was observed from the solar west limb 
on 2017 September 10. The eruption clearly shows the plasma sheet structure underneath the erupting flux rope 
predicted by magnetic reconnection models \citep[e.g.,][]{lin00, chen11}. 
The eruption was associated with an X8.2 flare 
(peak at 16:06 UT) from active region (AR)~12673 \footnote{\label{flarelist}https://www.lmsal.com/solarsoft/latest$\_$events/} 
observed by the Geostationary Operational Environmental Satellite (GOES). 
The AR~12673 produced four X-class flares from September 6. 
The interplanetary coronal mass ejections (ICMEs) associated with the X-class flares were observed in situ measurements. 
However, the ICME associated with the west limb flare on September 10 was observed in its flank, 
and the ICMEs from the preceding flares were detected in a merged structure \citep{guo18, bruno19, hajra20}. 
As a result, the in situ ion charge states are indistinguishable from those of preceding eruptions, 
thus preventing the application of freeze-in constraints from the ICME plasma. 

\begin{figure}
\centering
\includegraphics[width=150mm]{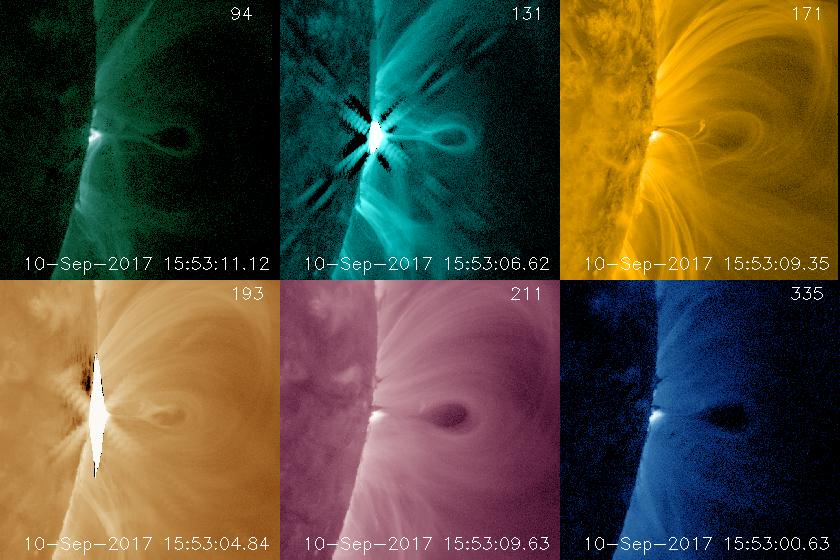}
\caption{AIA observations of a flux rope eruption on 2017 September 10. An animation of this figure is available.} 
\label{fig:aiafig}
\end{figure}

The AIA \citep{lemen11} on board SDO \citep{pesnell12} observes the solar corona and transition
region up to 0.5$R_\sun$ above the solar limb with high spatial
($\sim$ 0.6$''$/pixel) and temporal ($\sim$12 s) resolutions in seven
narrow EUV passbands (94 \AA, 131 \AA, 171\AA , 193\AA, 211 \AA,
304 \AA, 335 \AA). The flux rope eruption was observed in all EUV passbands. 
The flux rope erupts starting from $\sim$ 15:37 UT, and the plasma sheet structure underneath the flux rope was 
observed from around 15:50 UT and continued later than $\sim$ 20:00 UT (see Figure~\ref{fig:aiafig} and the accompanying animation).

\subsection{Tracking of a blob in the plasma sheet} \label{subsec:tracking} 

The AIA observations are calibrated prior to analysis. 
We use the standard point-spread function (PSF)\footnote{\label{psf}http://hesperia.gsfc.nasa.gov/ssw/sdo/aia/idl/psf/DOC/psfreport.pdf} 
to remove the diffraction and scattering effects from the
AIA observations. The PSF includes the pattern of these effects, calculated using the
procedure aia\_calc\_psf.pro at each passband of AIA.  We
deconvolve the Level 1.0 images with the point-spread function
using the procedure aia\_deconvolve\_richardsonlucy.pro. 
Then, we calibrate
the images to Level 1.5 using a standard SolarSoftWare (SSW) IDL procedure aia\_prep.pro that co-aligns and adjusts the plate scales and roll angles
between the AIA channels.

\begin{figure}
\centering
\includegraphics[width=150mm]{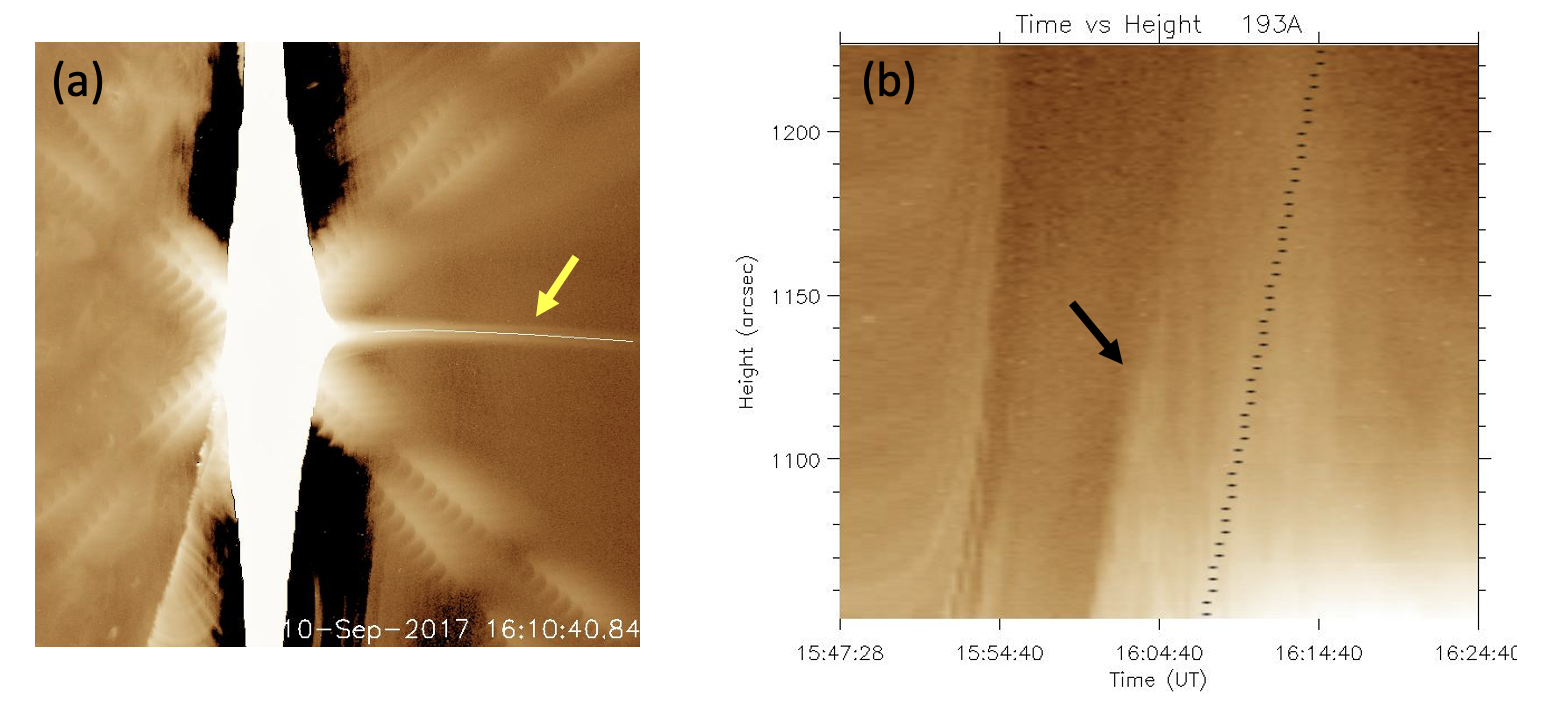}
\caption{(a) Slit placed along the plasma sheet at 16:10:40 UT in the 193 \AA\ channel (white solid line). 
The yellow arrow indicates the slit.
(b) Height information extracted along the black dotted line at each time for the slit coordinate boxes in Figure~\ref{fig:tracking}. 
The black arrow marks the onset of plasma sheet propagation.
}
\label{fig:ts193}
\end{figure}

\begin{figure}
\centering
\includegraphics[width=150mm]{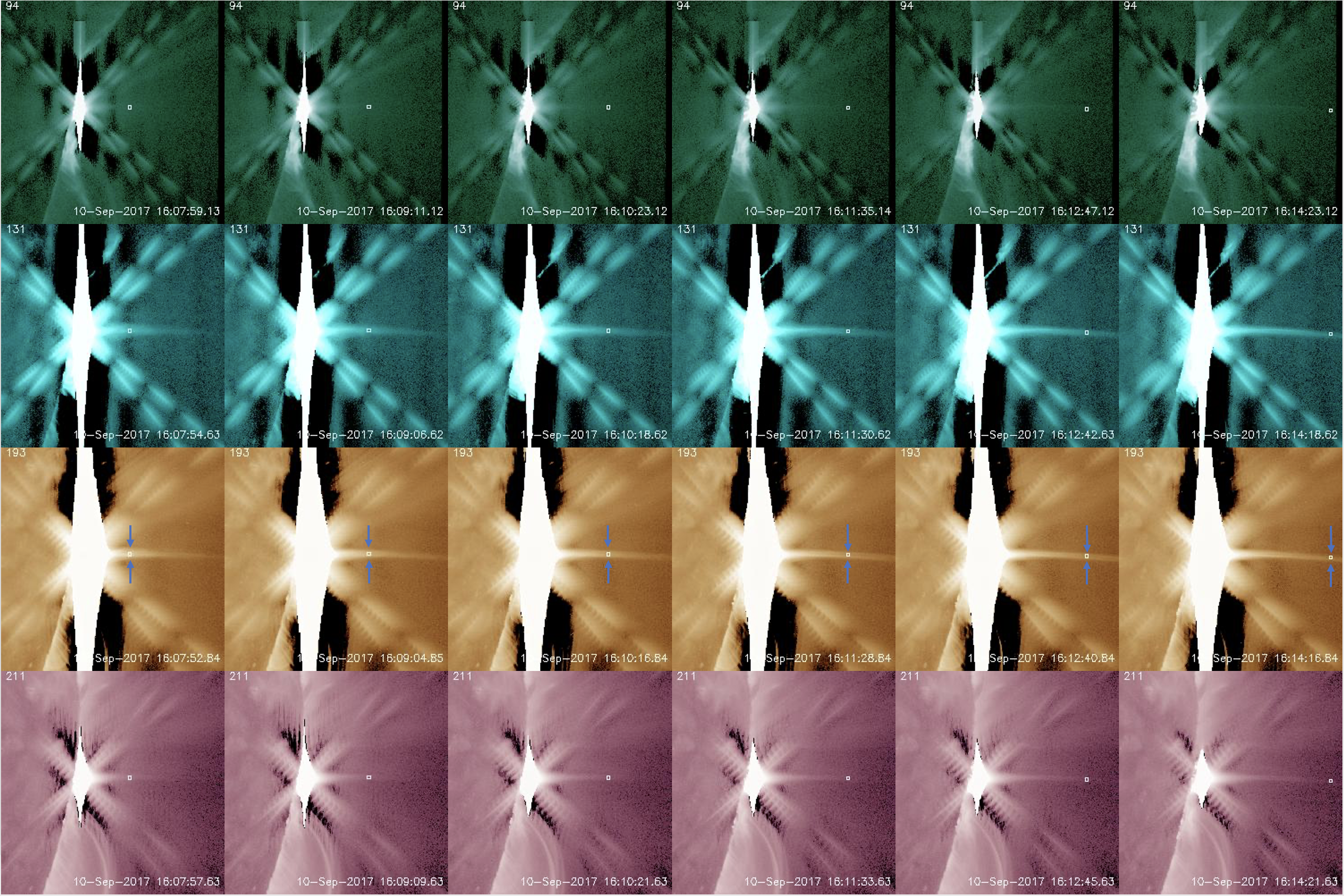}
\caption{Tracking of a blob within the plasma sheet in the 94, 131, 193, and 211 \AA\ channels (top to bottom). 
Blue arrows in the 193 \AA\ images indicate the locations where the line-of-sight depths are estimated. 
An animation of this figure is available.}
\label{fig:tracking}
\end{figure}

We track a blob in the plasma sheet using a time-space map. 
A slit is selected along the plasma sheet observed on 193 \AA\ at 16:10:40 UT(Figure~\ref{fig:ts193}~(a)), 
and the corresponding time-space map is constructed (Figure~\ref{fig:ts193}~(b)). 
The black arrow in Figure~\ref{fig:ts193}~(b) marks the onset of blob propagation along the plasma sheet. 
Because the intensity observed at the leading edge is too low, 
we instead track the blob along the black dotted line just behind the foremost front. 
The slope of this feature in the time-space map corresponds to an apparent speed of about $\sim$336~km~s$^{-1}$.
The coordinates of this dotted line provide the observed time-height information of the propagating blob. 
Using these pixel coordinates, we determine the blob locations at each time (Figure~\ref{fig:tracking}). 
Following this procedure, we track the blob from 16:07 UT to 16:14 UT. 
We use observations with exposure times longer than 2.9 sec in the 94 \AA, 131\AA, and 211 \AA\ and 1.9 sec for 193 \AA. 
The exposure times for the 171 \AA\ and 335 \AA\ are all  $\sim$2 sec and $\sim$2.9 sec, respectively. 
The tracking time is about 384~sec, corresponding to 17 AIA observations. 

\subsection{Densities of the plasma blob} \label{subsec:densities} 

\begin{figure}
\centering
\includegraphics[width=80mm]{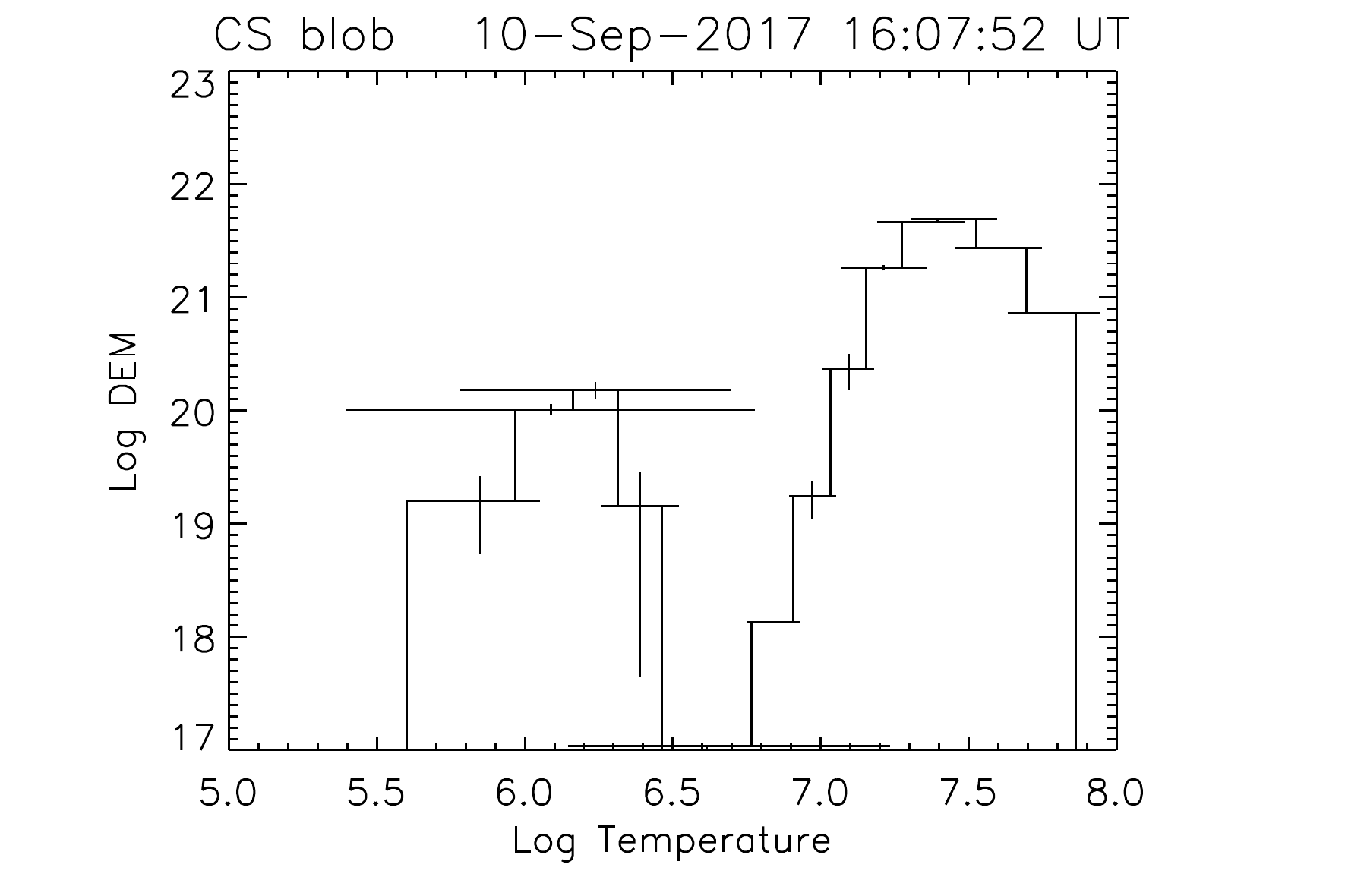}
\includegraphics[width=80mm]{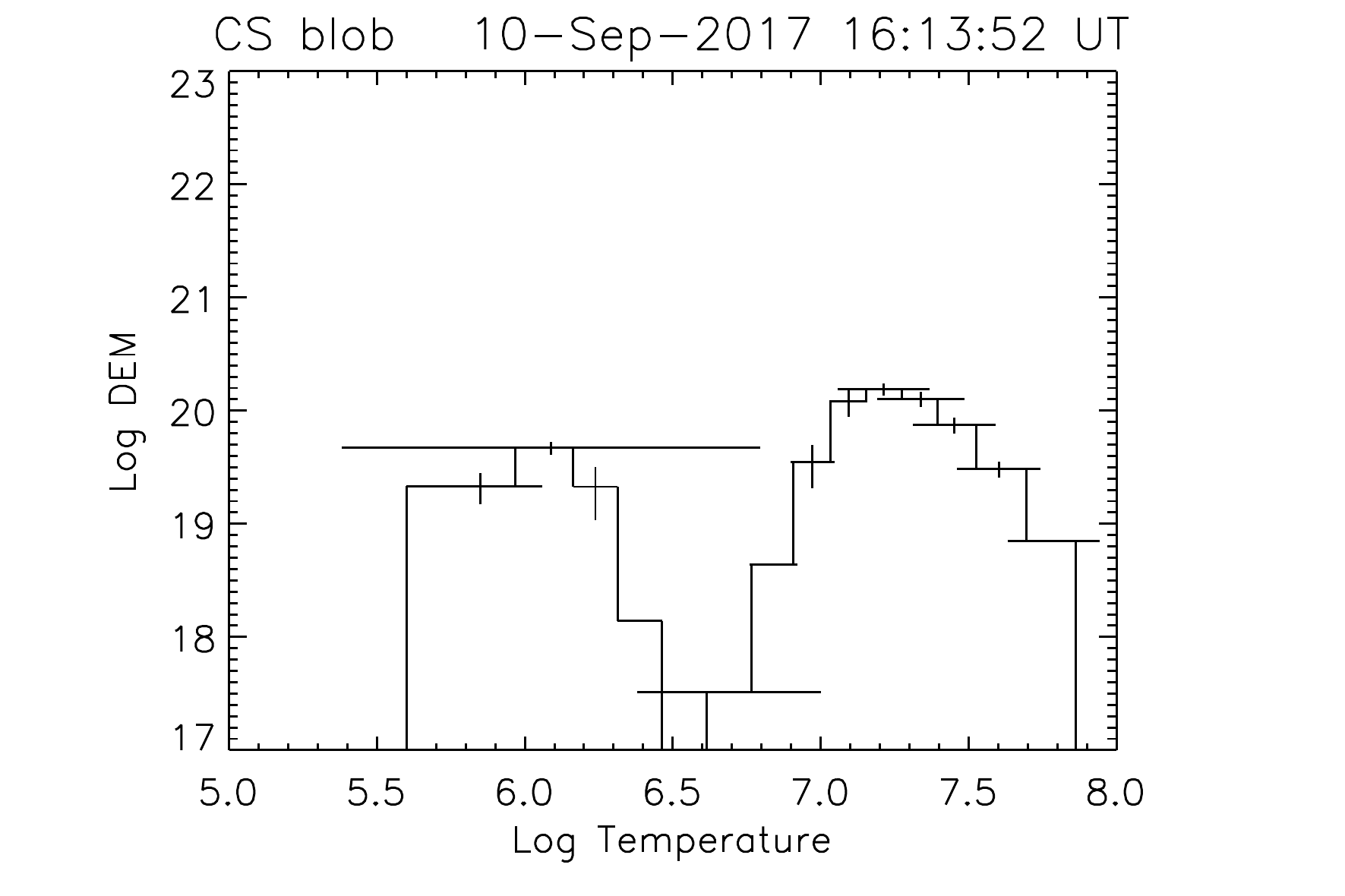}
\caption{DEMs at the initial and final times used to estimate the temperature and density of the tracked blob. Horizontal and vertical bars denote the uncertainties in temperature and DEM, respectively.}
\label{fig:dem}
\end{figure}

We find the densities of the plasma blob using a differential emission measure (DEM) method. 
We use a DEM code which performs a zeroth-order regularization to produce DEM maps and an optimized and faster version of a DEM code  developed by \citet{hannah12, hannah13}\footnote{\label{demreg}https://github.com/ianan/demreg}. This code has been used to determine the  temperature of an erupting flux rope \citep{lee17}. 
From these DEM results, we estimate the densities of the current sheet plasma blob, as well as the temperatures and densities of the background emission, for the pixels within the 10x10 pixel boxes shown in Figure~\ref{fig:dem}. 
The averages over these 100 pixels are used.  
For the plasma sheet blob, we model the temperatures with a grid of heating rates and impulsive heating scenarios (see Section 3) while we use the densities from these DEMs. 
We calculate the DEMs during about 384 seconds from 16:07 UT to 16:14 UT, corresponding to 17 observations with a 24-second cadence for long exposure observations. We show the DEMs for the first and the last times in Figure~\ref{fig:dem}. 
The DEMs show two bumps, indicating the background emission at a lower temperature and the plasma sheet blob at a higher temperature.
We calculate the emission measure using the equation.  

\begin{equation}
EM (t) = \sum_{j} DEM(t,j) * dT(j), 
\end{equation}

where EM({\it{t}}) is emission measure at each time step {\it t}, and {\it j} is the DEM temperature bin.  
For the plasma sheet blob, we sum the DEM over LogT = 6.9 to 7.8. 

We estimate the electron density by assuming that the plasma sheet has a cylindrical geometry, where the line of sight depth is taken to be equal to the observed thickness of the plasma sheet at 193 \AA\ at each time (see blue arrows in Figure~\ref{fig:tracking}). Then, we derive the densities using  
   
\begin{equation}
EM (t) = n_e(t)^2 dl(t), 
\end{equation}

where $n_e$ is the electron density and $dl(t)$ is the line of sight depth at each time. 

The background emission is obtained by summing the DEM over LogT = 5.8 - 6.6. 
The corresponding background temperature is derived as the EM-weighted temperature, as follows.

\begin{equation}
T (t) = \frac{\sum_{j} T(t,j)*DEM(t,j)} {EM(t)}.
\end{equation}

We show the line of sight depth, the plasma sheet density, and the background temperature and density in Figure~\ref{fig:density}. \citet{warren18} estimate the density of the plasma sheet for the same event as 10$^{10}$ cm$^{-3}$ at 16:41 UT using the AIA and EIS observations. \citet{gou24} find emission measures, 2.0$\times$10$^{27}\sim~$4.4$\times$10$^{25}$ cm$^{-5}$ and 2.2$\times$10$^{29}\sim~$1.0$\times$10$^{27}$   cm$^{-5}$, with the constant thickness, 7.1$\times$10$^8$ cm and 2.6$\times$10$^8$ cm, along the plasma sheet in the plane of the sky, at 15:55 UT and 16:41 UT, respectively. In comparison, we obtain a plasma sheet density of 8.8$\times$10$^9$ cm$^{-3}$ at 16:07 UT, assuming line of sight depths of 1.4$\times$10$^9$ cm. Although the observation times and locations are slightly different, our result is quantitatively very similar to the previous estimates by \citet{warren18} and \citet{gou24}.  For the background emission, we find that the temperature and density, $\sim$1.1~MK and $\sim$3.1$\times$10$^8$cm$^{-3}$ respectively, remain nearly constant over time. 

The derived density history of the plasma sheet, together with the background temperature and density histories, is used to construct the models in Section~\ref{sec:models}. 

\section{MODELS} \label{sec:models} 

We assume that the blob in the plasma sheet is rapidly heated to high temperatures by impulsive energy release during magnetic reconnection. After this initial impulsive heating, the blob propagates outward while undergoing adiabatic expansion, which alone would lead to a much faster temperature decline than indicated by the observations. We therefore include an additional continuous heating term during the propagation. Our analysis begins when the blob becomes clearly identifiable in the AIA, which occurs after the onset of the flare and the initial development of the current sheet. Therefore, the impulsive heating term in our model does not necessarily correspond to the timing of the global flare peak, but rather represents an effective, localized energization event that sets the initial thermodynamic state of the tracked blob at the start time. 

We construct the time-dependent temperature histories of the current sheet plasma, and apply the temperature and density histories to a time-dependent ionization to compute the ion fractions in the presence of nonthermal electrons. The ion fractions are used to synthesize the AIA observations, which are compared with the uniquely well-observed plasma sheet. This allows us to understand the physical conditions experienced during the plasma propagation. In this section, we describe how we model the time-dependent temperature evolution, compute the ion fractions in NEI with nonthermal electrons, and compare the synthesized emission with the observations. 

\subsection{Temperature history of the plasma sheet} \label{subsec:temperature}

We construct time-dependent temperature histories of the plasma sheet by parameterizing the peak temperatures produced by the impulsive heating and continuous heating rates, while accounting for cooling due to adiabatic expansion. All models begin with a background temperature of 1.3~MK at the first observation time (16:07 UT), derived in Section~\ref{subsec:densities}, and are then heated impulsively to the peak temperatures. For the impulsive heating, we adopt the peak temperatures (hereafter HT) ranging from 2~MK to 50 MK (logT=6.3$\sim$LogT=7.7 with an interval of $\Delta$ logT=0.1). For the continuous heating, we adopt heating rates (hereafter HR) of 0.02 to 2 erg cm$^{-3}$ s$^{-1}$ (0.02, 0.04, 0.08, 0.12, 0.16, 0.2, 0.4, 0.8, 1.2, 1.6, 2.0 erg cm$^{-3}$ s$^{-1}$), accounting for temporal decreases of 10$\%$, 66$\%$, and 80$\%$ as the blob propagate along the plasma sheet. 

Therefore, we obtain a total of 495 heating models consisting of 15 models for impulsive heating and 33 models for HR (11 HR models $\times$ 3 reduction factors). Including the 13 $\kappa$ values, which represent the nonthermal electron velocity distributions, described in Section 3.2, the final number of models becomes 6435. 

We obtain the temperature history by adding the continuous HRs to the temperature evolution that results from impulsive heating followed by adiabatic expansion cooling, as given in the following equation: 

\begin{equation}
\label{eq:Tcs}
T_{cs}(t) = T_{ad}(t) + \Delta T_{ch}(t)
\end{equation}

\begin{figure}
\centering
\includegraphics[width=50mm]{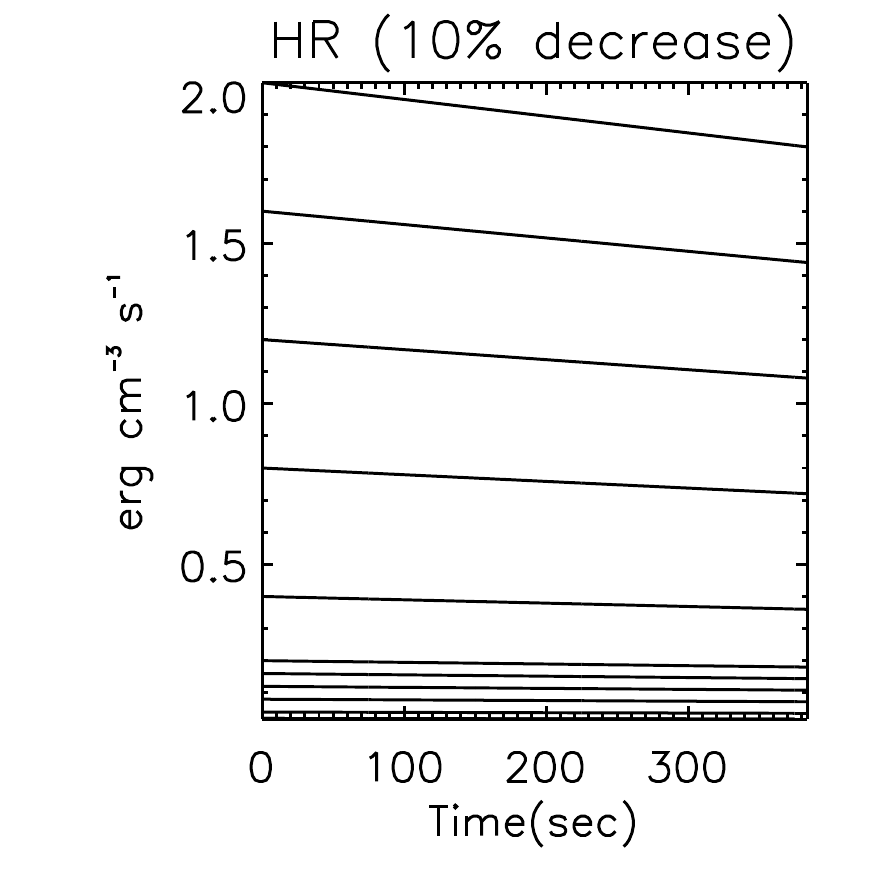}
\includegraphics[width=50mm]{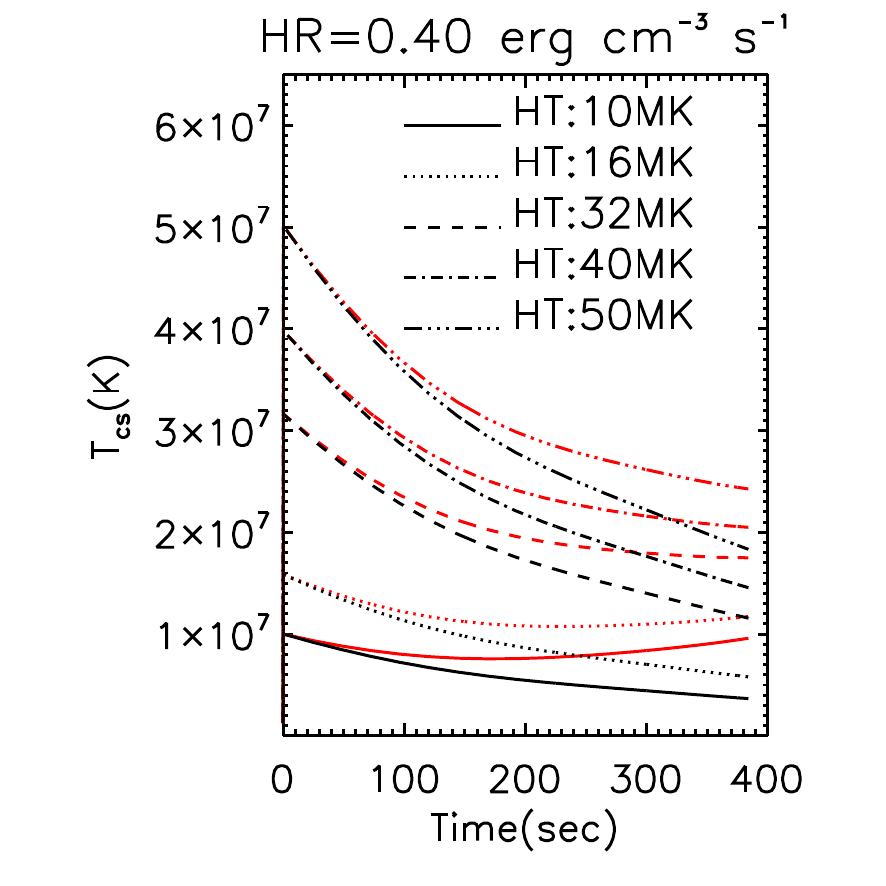}
\includegraphics[width=50mm]{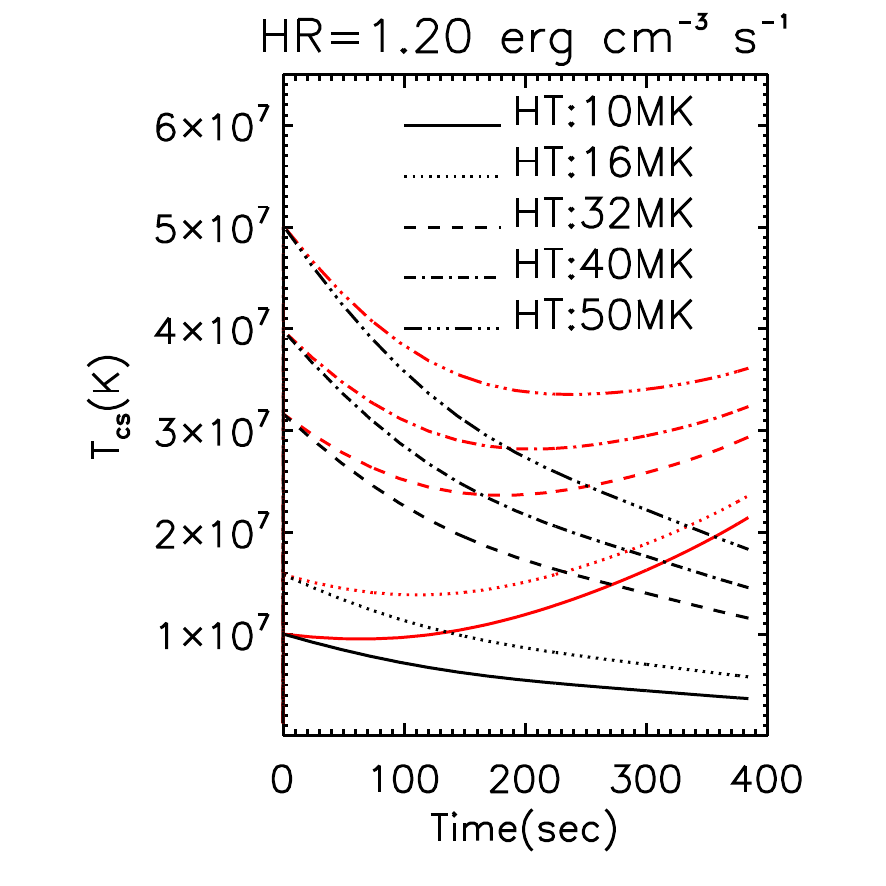}
\includegraphics[width=50mm]{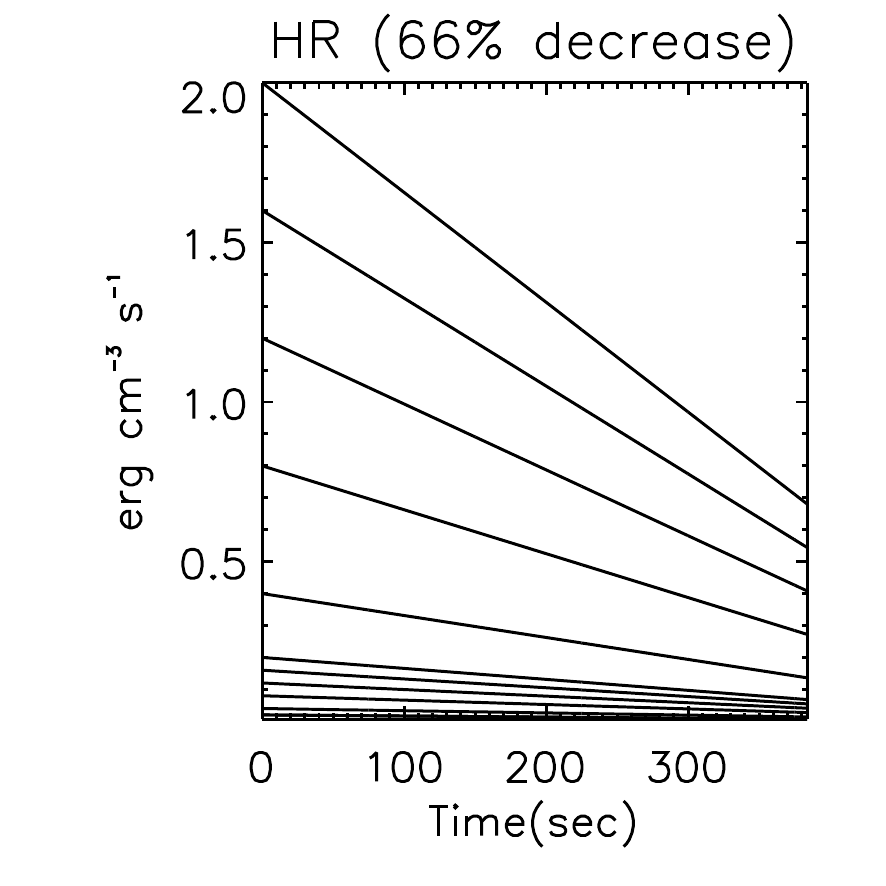}
\includegraphics[width=50mm]{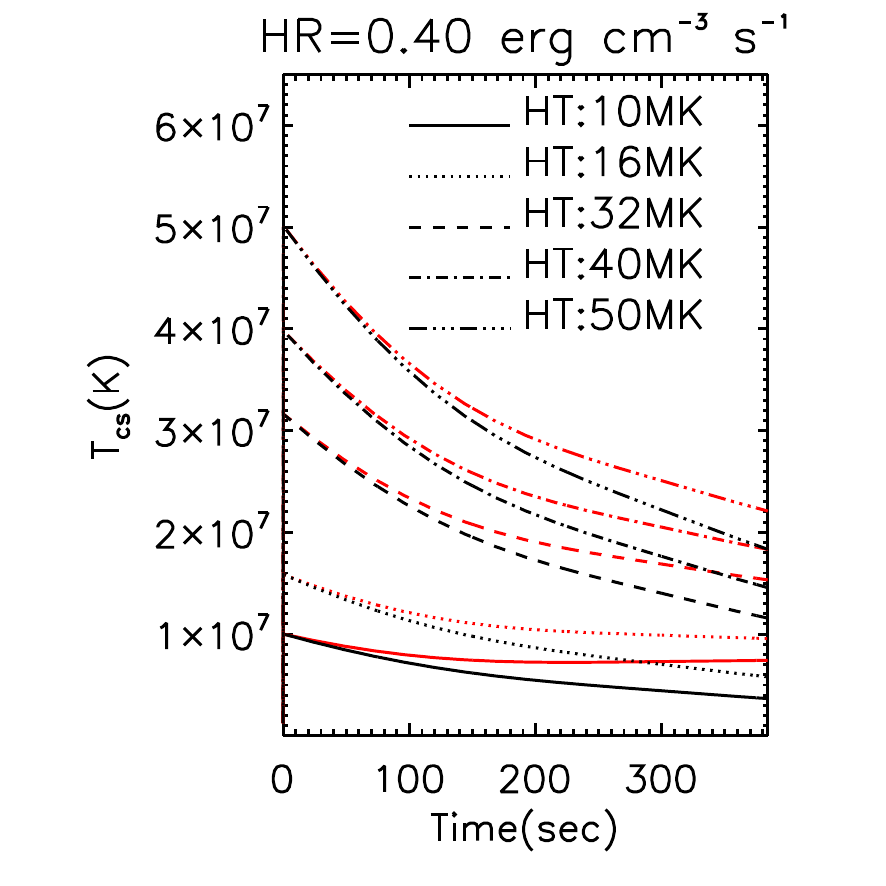}
\includegraphics[width=50mm]{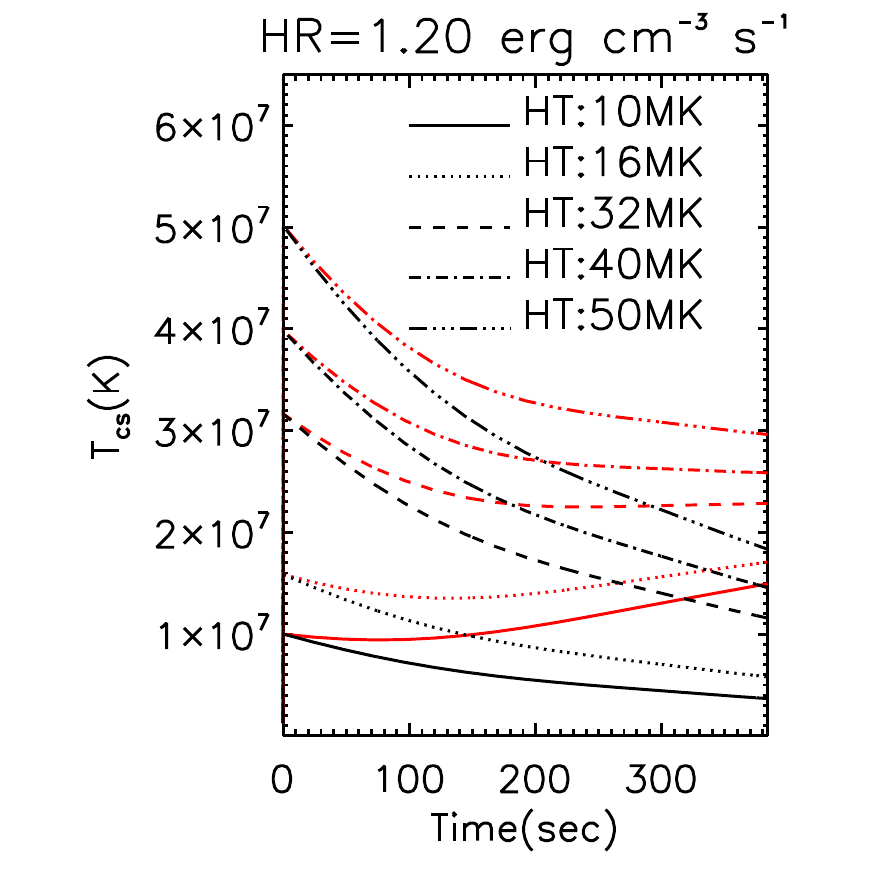}
\includegraphics[width=50mm]{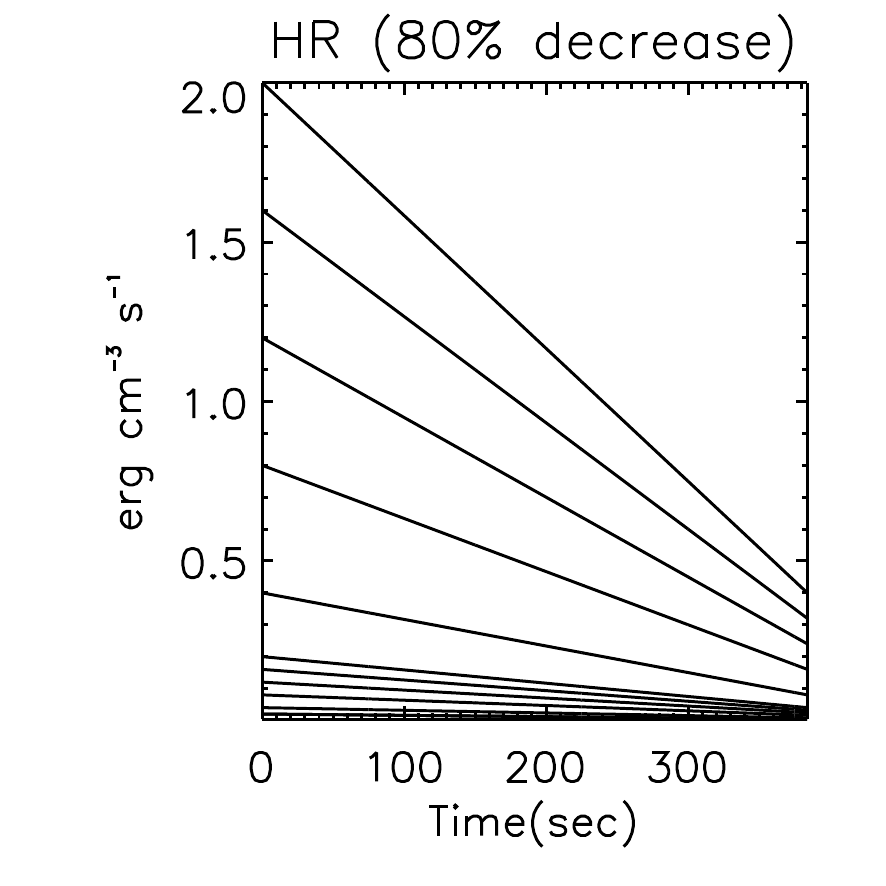}
\includegraphics[width=50mm]{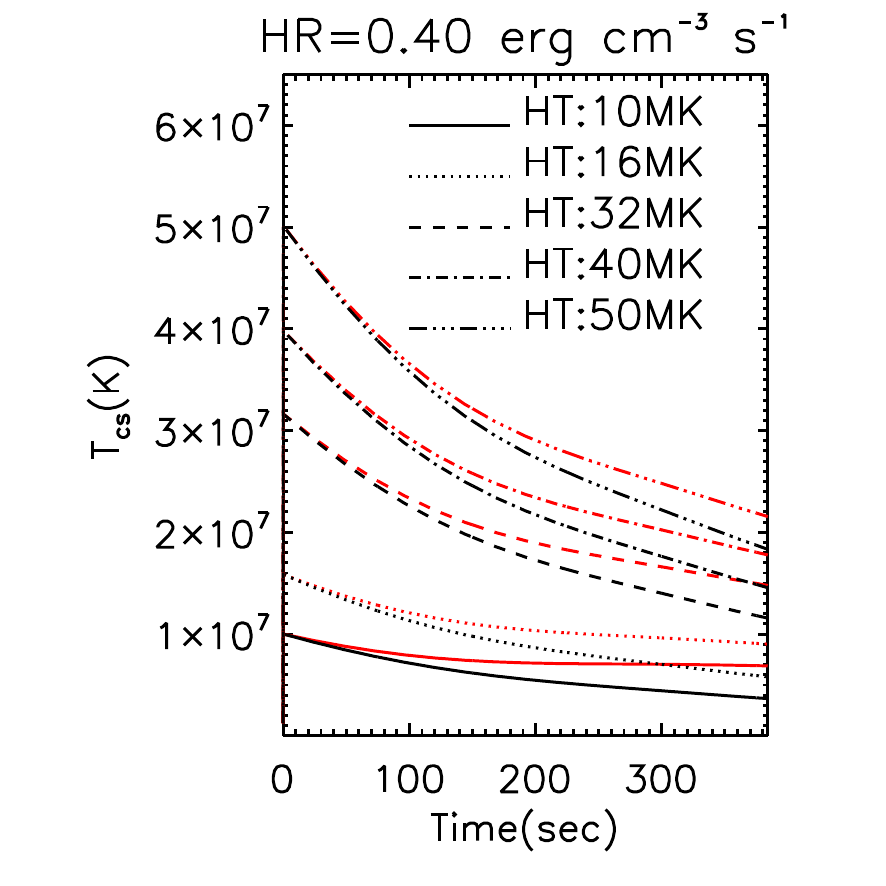}
\includegraphics[width=50mm]{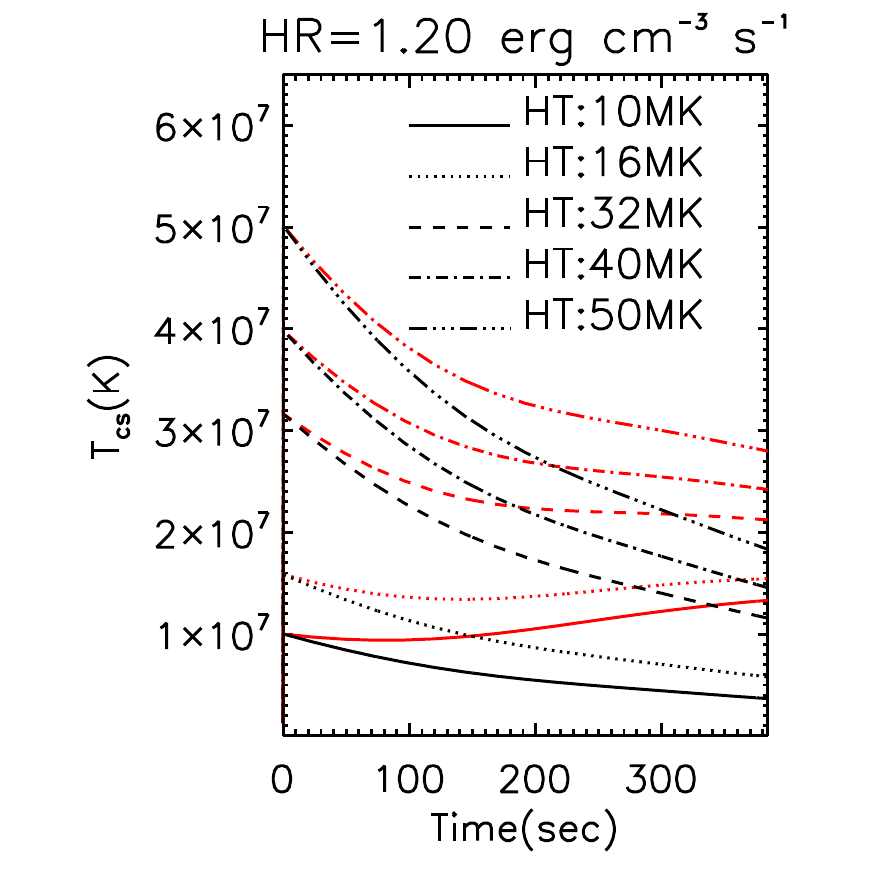}
\caption{Temperature histories for several combinations of $HT$ and $HR$. The top, middle, and bottom panels correspond to 10\%, 66\%, and 80\% reductions, respectively. 
Black and red curves denote cases without and with continuous heating, respectively. 
The first column shows varying HRs, while the second and third columns present 
$T_{\rm CS}$ (Equation~\ref{eq:Tcs}) for HR = 0.4 and 1.2 erg cm$^{-3}$ s$^{-1}$, respectively.}
\label{fig:heating}
\end{figure}

Here, $T_{cs}(t)$ is the temperature history of the plasma sheet, $T_{ad}(t)$ is the temperature evolution due to adiabatic expansion cooling after the impulsive heating, $\Delta T_{ch}(t)$ is the temperature increase due to continuous heating, and $t$ is time. Radiative cooling can be neglected for densities $<10^{10} \mathrm{cm^{-3}}$ and $T>10^7$ K for the plasma blob we follow. \citet{Kittrell25} showed that radiative cooling is negligible relative to conductive cooling in the plasma sheet during the 2017 September 10 event. The two terms are given by, 

\begin{equation}
\label{eq:Tad}
T_{ad}(t) = T{(1)} \left(\frac{n_e(t)}{n_{e}(1)}\right)^{(\gamma-1)}, 
\end{equation}

and

\begin{equation}
\label{eq:Tch}
\Delta T_{ch}(t) = \sum_{t} \frac{2}{5 k_{B}} \frac {H_r(t)} {n(t)} dt, 
\end{equation}

where $T(1)$ is the HT produced by the impulsive heating, $n_e(1)$ is the electron density at the time of the impulsive heating, and $n_e(t)$ is the time-dependent electron density described in Section~\ref{subsec:densities}. The adiabatic index is set to $\gamma$=5/3, and we use a time step of $dt$=0.1 s. The temperature increment $\Delta T_{\rm ch}$ due to continuous heating is calculated using an energy equation. $H_r(t)$ represents the continuous HR. We apply reductions of 10$\%$, 66$\%$, and 80$\%$ to the HR over the entire tracking duration of 384~s. $k_{\rm B}$ is Boltzmann constant. $n(t)$ is the plasma number density, assuming that the electron and proton densities are the same. The continuous HR includes all contributions to the heating, such as wave dissipation, magnetic reconnection, turbulent heating, or the divergence of conductive flux, or shock waves generated by the reconnection outflow. We show the temperature histories for several heating models in Figure~\ref{fig:heating}. 
 
\subsection{Ion fractions in NEI with nonthermal electrons}

Once the time-dependent temperature histories are obtained from the models in Section~\ref{subsec:temperature}, we apply them, together with the density histories estimated in Section~\ref{subsec:densities}, to a time-dependent ionization model \citep{shen15} to compute the ion fractions. We note that the density histories in Section~\ref{subsec:densities} were derived under the assumption of ionization equilibrium, whereas the temperature histories are modeled in Section~\ref{subsec:temperature}. 
While the rate coefficients are functions of temperature and $\kappa$, 
the electron density affects both the evolution timescale and the emission scale. 
Because the same density history is used for all models, 
the relative differences among cases with different HT, $\kappa$, and heating rates remain robust.

NEI has a significant effect on the relative count rates in the different bands,
but it primarily shifts the charge states by one or two ionization stages, and therefore 
redistributes emission from one band to another. This leads to only a modest effect on the EM, and therefore on the
density, which is the quantity used in this study. The resulting effect is expected to be $\sim$20\% effect, comparable to the
uncertainty due to geometric assumption used to derive density from EM. 
\citet{shi2019} show that NEI can change spectral line intensity by up to 120\%, 
while the inferred density is not significantly affected, with differences of about 10\% at height below 1.1 R$_\sun$ 
when line-of-sight integration is considered. This supports our use of density derived under the equilibrium assumption for relative model comparisons.
 
We calculate the ion fractions for all charge states of the 16 most abundant elements, H, He, C, N, O, Ne, Na, Mg, Al, Si, S, Ar, Ca, Cr, Fe, and Ni. The ionization model pre-computes the ionization and recombination rates, and the rates are saved into tables containing the eigenvalues and eigenvectors for fast calculation. This time-dependent ionization model has previously been used to investigate the NEI effects in rapidly heated plasma and their impact on solar EUV and X-ray imaging observations \citep{lee19}.  

In this work, we incorporate nonthermal electron velocity distributions into the ionization model to account for departures from thermal equilibrium caused by particle acceleration. Specifically, we adopt $\kappa$ distributions, characterized by a Maxwellian core and suprathermal power-law tails. The $\kappa$ distributions have been widely used to describe space and astrophysical plasmas in which wave-particle interactions or acceleration processes generate enhanced high energy populations\citep[e.g.,][]{livadiotis13}. 

We pre-compute the ionization and recombination rates for various $\kappa$ distributions using the Maxwellian decomposition method\citep{cui19} and CHIANTI~9\citep{dere19}\footnote{\label{nonMax}https://github.com/ionizationcalc/Non-MaxwellianDistribution}. The rates are calculated for 13 values of ~$\kappa$($\kappa=$2, 3, 4, 5, 6, 8, 10, 12, 15, 20, 25, 30, 100) and saved into tables. Using the pre-computed tables, the ion fractions are calculated with the time-dependent ionization equation, 

\begin{equation}
\label{eq:ionization}
\frac {df_{i}}{dt} = n_{e} [C_{i-1} f_{i-1} - (C_{i} + R_{i})f_{i} + R_{i+1}f_{i+1}]
\end{equation}

\noindent where $f_i$ is ion fraction with charge state $i$, $n_e$ is electron density, and $t$ is time. The coefficients $C_i$ and $R_i$ are ionization and recombination rate coefficients, which depend on the electron temperature and the chosen $\kappa$ values. We use the electron temperatures (T$_{cs}$) of the plasma sheet from Equation~\ref{eq:Tcs}.
 
\subsection{Synthesis of AIA observations from the models} 

We synthesize the AIA observations using the ion fractions computed with Equation~\ref{eq:ionization}, together with the various impulsive HTs, T(1) in Equation~\ref{eq:Tad}, continuous HRs in Equation~\ref{eq:Tch}, and $\kappa$ values. The synthesized digital numbers (DN$_{syn}$[DN~s$^{-1}$]) are then compared with the uniquely well-observed plasma sheet (DN$_{obs}$[DN~s$^{-1}$]). We compare the DN$_{syn}$ in 94\AA, 131\AA, 193\AA, and 211\AA\ bands with the corresponding AIA observations, as these channels clearly reveal the plasma sheet structure. 
The plasma sheet emission in the 94~\AA\ channel is considerably fainter than in the 131~\AA, 193~\AA, and 211~\AA\ channels, although the structure remains detectable. The cooler 171~\AA, 335~\AA, and 304~\AA\ channels do not clearly show the plasma sheet and are therefore not included in the analysis.

The AIA observations contain both the emission from the background and the blob of plasma sheet along the line of sight. Therefore, we compare the observations with the DN$_{syn}$ that is given by 

\begin{equation}
DN_{\rm syn}(t, \mathrm{band}) = DN_{\rm bg}(t, \mathrm{band}) + DN_{\rm cs}(t, \mathrm{band}), 
\end{equation}

\noindent where DN$_{bg}$(t, band) and DN$_{cs}$(t, band) are the background emission and plasma sheet blob emission at each AIA bandpass, respectively, defined as 

\begin{equation}
DN_{\rm bg}(t, \mathrm{band}) = R_{\rm eq} (T_{\rm bg}(t), \mathrm{band}) \times \mathrm{EM_{\rm bg}(t)}, 
\end{equation}

\begin{equation}
DN_{\rm cs}(t, \mathrm{band}) = R_{\rm neq}(T_{\rm cs}(t), \kappa, \mathrm{band}) \times \mathrm{EM_{\rm cs}(t)}. 
\end{equation}

\begin{figure}
\centering
\includegraphics[width=80mm]{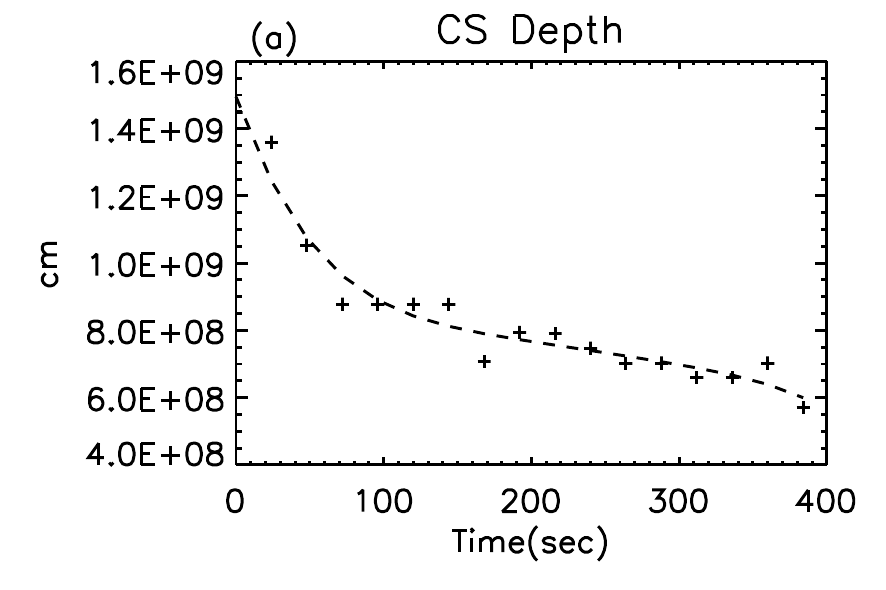}
\includegraphics[width=80mm]{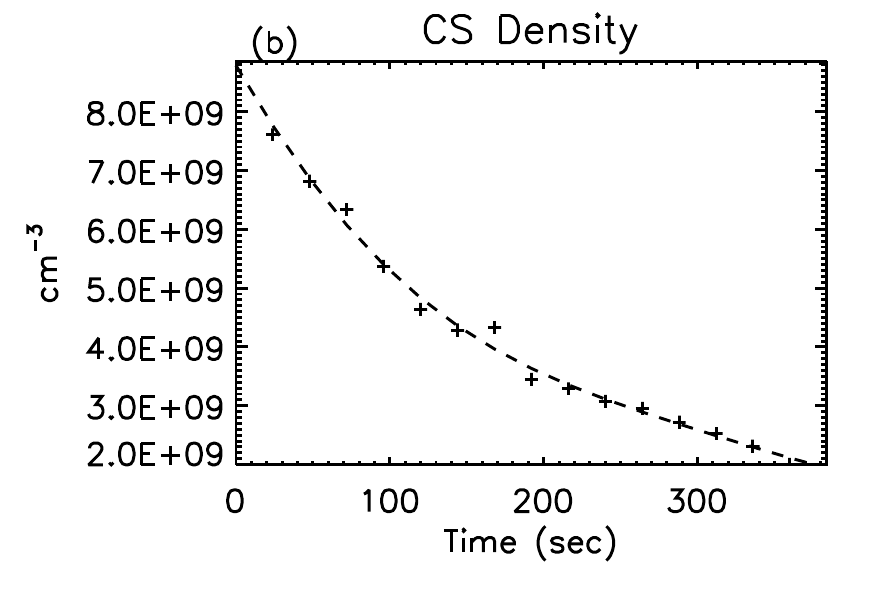}
\includegraphics[width=80mm]{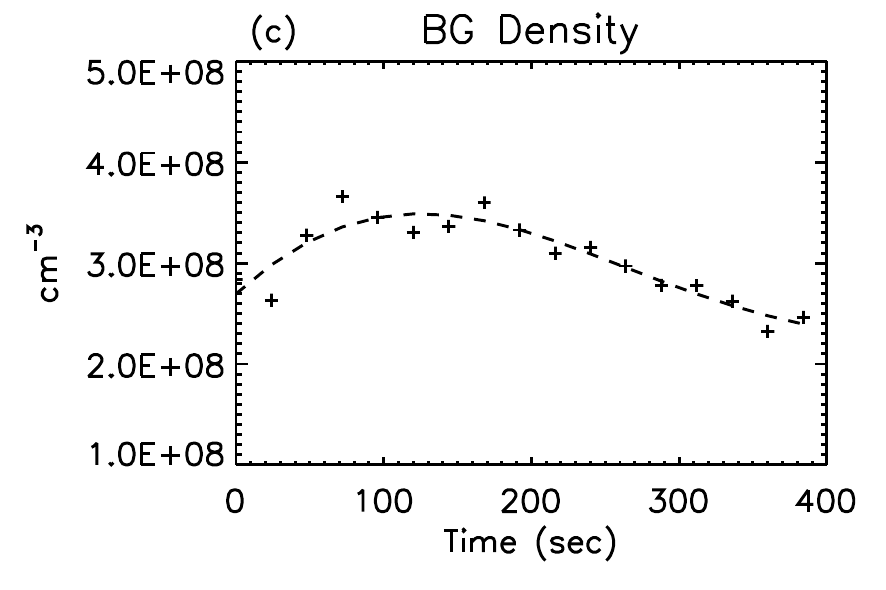}
\includegraphics[width=80mm]{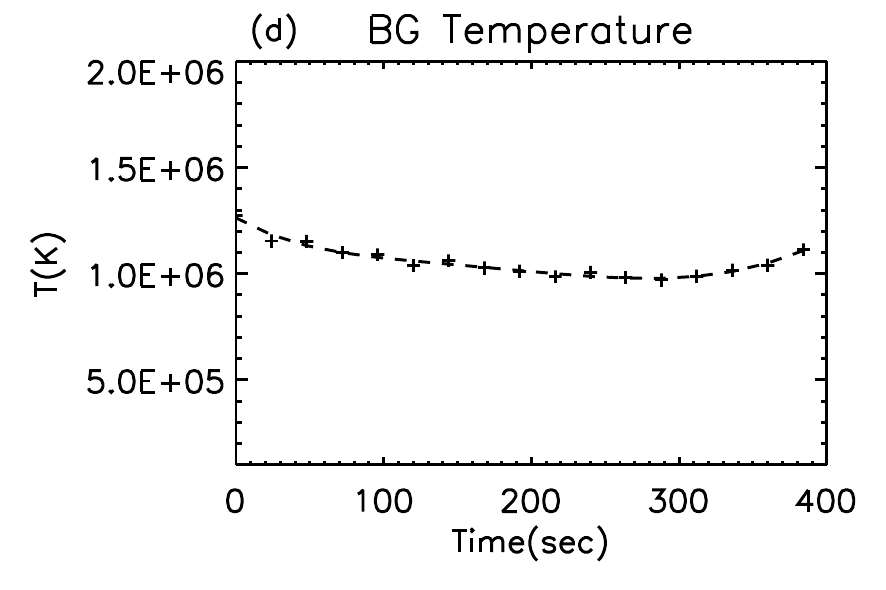}
\caption{Line-of-sight depth (a) and electron density (b) of the plasma sheet, and electron density (c) and temperature (d) of the background emission.}
\label{fig:density}
\end{figure}

\noindent Here R$_{eq}$ is the AIA temperature response at the background temperature T$_{bg}(t)$, calculated in Section~\ref{subsec:densities} (see Figure~\ref{fig:density}(d)) under the assumption of ionization equilibrium in SSW. R$_{neq}(T_{cs}(t),\kappa, band)$ is the nonequilibrium temperature response defined in Equation~\ref{eq:rneq} calculated using the ion fractions from  Equation~\ref{eq:ionization}, together with the modeled plasma sheet temperatures $T_{cs}(t)$ from Equation~\ref{eq:Tcs} and the adopted $\kappa$ distributions. 
The quantities EM$_{bg}$(t) and EM$_{cs}$(t) are the emission measure of the background and plasma sheet blob, respectively, defined as  

\begin{equation}
EM_{\rm bg}(t) = \langle n_{e,{\rm bg}(t)}^2 \rangle dl(t), 
\end{equation}

\begin{equation}
EM_{\rm cs}(t) = \langle n_{e,{\rm cs}(t)}^2 \rangle dl(t), 
\end{equation}

\noindent where n$_{e,bg}$(t) and n$_{e,cs}$(t) are the electron densities of the background and plasma sheet, respectively, and $dl$(t) is the line of sight depth calculated in Section~\ref{subsec:densities} (Figure~\ref{fig:density}(a)). 

The nonequilibrium temperature response is calculated using the same formulation as in \citet{lee19}, but extended to include the $\kappa$ distributions, as expressed by 

\begin{equation}
\label{eq:rneq}
R_{\rm neq}(T_{\rm cs}(t), \kappa, \mathrm{band}) = \sum_{Z} \sum_{z} Resp(Z,z,T_{\rm cs}(t),\kappa, \mathrm{band}) AB(Z) f(Z,z,T_{\rm cs}(t),\kappa) + \sum_{\lambda} \mathrm{cont}(T_{\rm cs}(t), \lambda) A_{\rm eff, \mathrm{band}} (\lambda).
\end{equation}

The continuum, $cont(T_{cs}, \lambda)$, is computed using the SSW routines 
\texttt{freefree.pro}, \texttt{freebound.pro}, and \texttt{two\_photon.pro}, 
which evaluate the bremsstrahlung, free-bound, and two-photon emission components, respectively.
The term Resp(Z, z, T$_{cs}$(t), $\kappa$, band)~[DN~cm$^5$~s$^{-1}$] is the temperature response for element Z, charge state z, temperature (T$_{cs}$(t)), and the AIA passband (band), and is defined as 

\begin{equation}
\label{eq:emissivity}
Resp(Z,z,T, \kappa, \mathrm{band}) = \sum_{i=1}^{\rm nlines} \frac {\epsilon{(i, T)}}{10^9} A_{\rm eff, \mathrm{band}} (\lambda (i)), 
\end{equation}

We pre-calculate the temperature response function in Equation~\ref{eq:emissivity} over the temperature (T) range from 10$^5$~K to 10$^8$~K, and then interpolate the values corresponding to the modeled plasma sheet temperature T$_{cs}$(t). 

The line emissivity, $\epsilon(i, T)$ [photon~s$^{-1}$], is calculated at an arbitrary density, $n_e$=10$^9~$cm$^{-3}$ using a procedure (emiss\_calc.pro) in CHIANTI~9 \citep{dere19}, and is then divided by the density. Because photoexcitation and stimulated emission are not included, the density is independent in the ionization balance and does not affect the temperature responses. We also include the transitions produced by dielectronic recombination. 

The effective area, A$_{eff,band}(\lambda)$~[cm$^2$ DN photon$^{-1}$] represents the wavelength-dependent sensitivity of the AIA instrument for each passband as a function of the wavelength ($\lambda$) for each transition line ($i$). We obtain A$_{eff,band}(\lambda)$ using a procedure, aia\_get\_response.pro, in SSW for 2017 September 10, accounting for the time-varing degradation of the instruments \citep{boerner2014, narukage2011}.  

The element abundance $AB(Z)$ is taken from a coronal abundance (sun\_coronal\_1992\_feldman\_ext.abund) in CHIANTI \citep{feldman92, landi02, grevesse98}.  The ion fraction f(Z,z,T$_{cs}$(t),$\kappa$) is derived from the time-dependent ionization model (Equation~\ref{eq:ionization}), which represents both nonequilibrium ionization and nonthermal electron distributions. 
Finally, we find the nonequilibrium temperature response R$_{neq}$ in units of DN~cm$^5$s$^{-1}$pix$^{-1}$ multiplying by $\Omega$/4$\pi$, where $\Omega$ is given by the pixel size, 0.6$''$.

\subsection{Model selection using covariance-weighted \texorpdfstring{$\chi^2$}{chi2}}

We search for the model that best matches the AIA observations among the various impulsive HTs, continuous HRs, and $\kappa$ distributions. At earlier times, rapid temperature variations driven by impulsive heating lead to correspondingly rapid changes in the ion fractions, making it difficult for the models to reproduce the observations. Therefore, we perform the comparison at times later than 48~s, excluding the first two observed points. In addition, the 94 \AA\ channel has relatively low signal-to-noise ratios, and its late-time values are not reliable. Accordingly, the comparison is restricted to the time range from 48~s to 382~s. We compare the models with 14 observed time points in four EUV passbands.

To identify the best nonequilibrium model among a large grid of NEI models with nonthermal electrons, 
we quantitatively compare the synthesized AIA light curves (DN$_{syn}$) with the observations (DN$_{obs}$) 
using a covariance weighted $\chi^2$ metric, 

\begin{equation}
\label{eq:chisq_compact}
\chi^2 = \mathbf{r}^{\mathrm T}\,\mathbf{C}^{-1}\,\mathbf{r},
\end{equation}

where the residual vector is defined as

\begin{equation}
\label{eq:residual_vector}
r(\mathrm{band},t) = DN_{\mathrm{obs}}(\mathrm{band},t) - a_{\mathrm{band}}\,DN_{\mathrm{syn}}(\mathrm{band},t).
\end{equation} 

All sources of uncertainty enter the goodness-of-fit metric through the covariance matrix $\mathbf{C}$, which we decompose as
\begin{equation}
\mathbf{C} = \mathbf{C}_{\rm stat} + \mathbf{C}_{\rm corr}.
\end{equation}

The statistical component $\mathbf{C}_{\rm stat}$ is derived using the procedure \textit{aia\_bp\_estimate\_error.pro} in SSW, configured to return only the statistical uncertainty component. This term includes photon noise and instrumental or processing-related contributions (readout, dark current, quantization, compression, and camera gain), and is treated as independent between passbands and time samples.

The correlated component $\mathbf{C}_{\rm corr}$ accounts for uncertainties that are correlated across EUV passbands. The same procedure also provides an uncertainty estimate using the \textit{evenorm} and \textit{temperature} keywords, which includes both statistical and additional systematic contributions associated with CHIANTI atomic data and calibration effects. The non-statistical portion of this total uncertainty is used to construct a correlated systematic component within the covariance matrix.

In our implementation, $\mathbf{C}_{\rm corr}$ can be written schematically as
\begin{equation}
\mathbf{C}_{\rm corr} = \mathbf{C}_{\rm band} + \mathbf{C}_{\rm sys},
\end{equation}
where $\mathbf{C}_{\rm band}$ represents a signal-proportional band-correlated term, and $\mathbf{C}_{\rm sys}$ corresponds to the correlated systematic component associated with CHIANTI atomic data and calibration-related uncertainties.
The term $\mathbf{C}_{\rm band}$ models amplitude variations shared across EUV passbands and is parameterized by a fractional amplitude $g_0$ and an inter-band correlation coefficient $\rho_{\rm band}$. The overall amplitude of $\mathbf{C}_{\rm sys}$ is scaled by $\alpha_{\rm sys}$, and its inter-band correlation structure is governed by $\rho_{\rm sys}$. In both cases, the correlated terms act only at the same time sample (time-diagonal approximation).

To account for residual amplitude mismatches among the AIA passbands, we allow band-dependent normalization factors $a_{\mathrm{band}}$ in Eq.~(\ref{eq:residual_vector}). The normalization factors are kept close to unity so that amplitude scaling does not artificially compensate for differences in the physical model parameters. The adopted covariance parameters and the constraint width $\sigma_{ab}$ for the band normalization factors $a_{\mathrm{band}}$ are summarized in Table~\ref{tab:dpar}.

Specifically, the AIA temperature response functions at high temperatures are subject to uncertainties in atomic data and calibration.
The adopted CHIANTI uncertainties of 50\% for the 94 and 131 \AA\ channels and 25\% for the 193 and 211 \AA\ channels, as in SSW, are included in the covariance matrix.  At 
high temperatures, the response may be dominated by bremsstrahlung continuum
rather than by the nominal emission lines, so the count rate is only weakly dependent
on temperature. These uncertainties do not significantly affect the inferred temperatures, as the constraints are primarily driven by the relative temporal evolution across multiple channels.

Finally, models with different HR decay scenarios (10\%, 66\%, and 80\%) are evaluated using the same fitting approach. The resulting parameters for representative models and their physical implications are presented in the following section.

\section{Results}

We investigate how the EUV observations constrain impulsive HT, continuous HR, and the nonthermal parameter $\kappa$ under three different HR decay scenarios. Using the covariance-based $\chi^2$ analysis, we first analyze the global $\chi^2$ behavior across the model grid, then characterize the structure of the confidence region. We then investigate how these constraints inform our interpretation of the ionization evolution and the heating mechanisms operating within the reconnecting current sheet.

\subsection{$\chi^2$ Variations} 

\begin{figure}
\centering
\includegraphics[width=80mm]{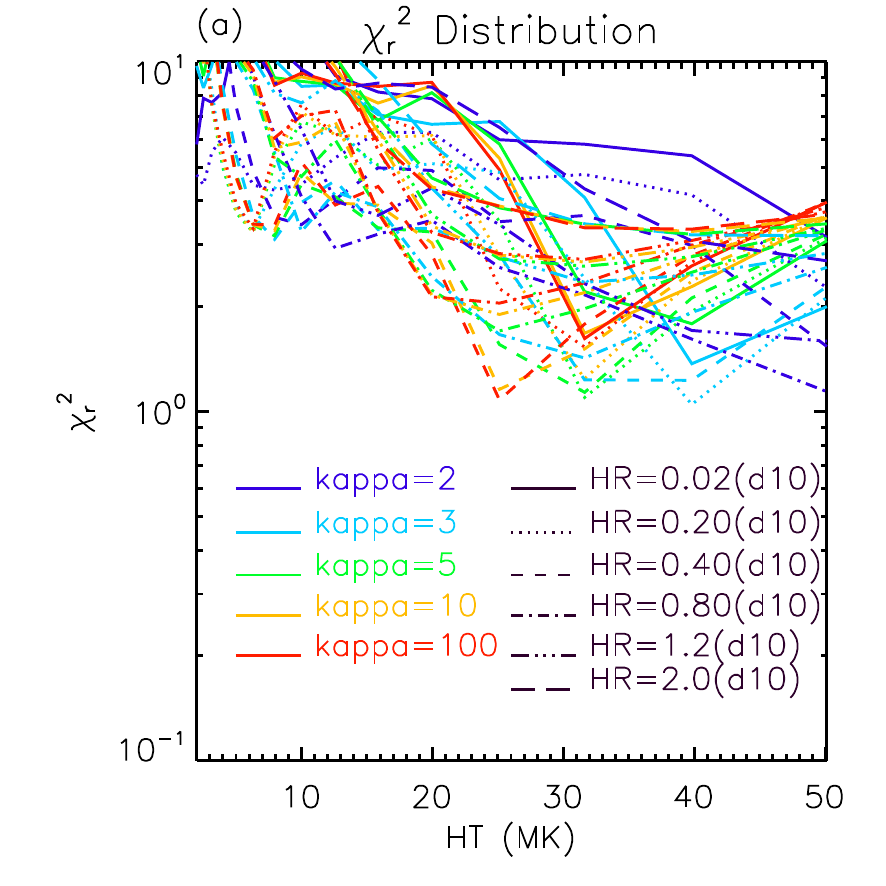}
\includegraphics[width=80mm]{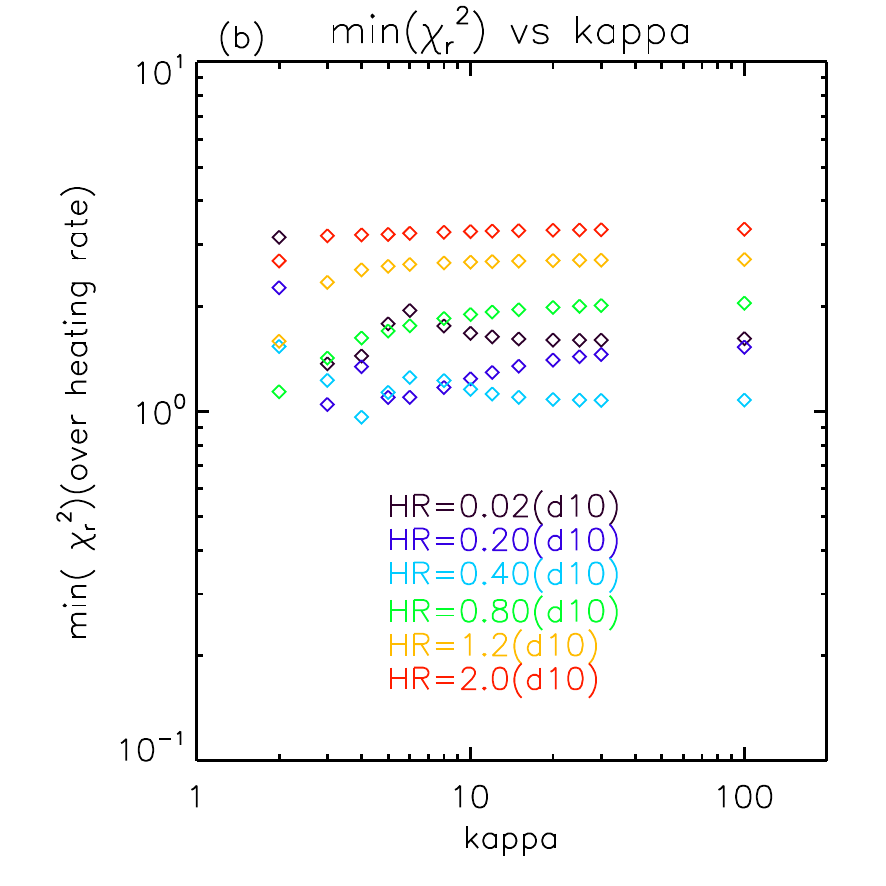}
\caption{(a) Distribution of $\chi_r^2$ for different model combinations. Colors and line styles represent selected values of $\kappa$ and HR, respectively.
(b) Minimum $\chi_r^2(\kappa)$ for each selected HR, obtained by minimizing over HT. Different colors correspond to different HR values.}
\label{fig:chi}
\end{figure}

We examine how $\chi^2$ varies across the explored parameter space. 
Figure~\ref{fig:chi}(a) shows the distribution of reduced $\chi^2$ ($\chi_r^2$) values 
for six HRs and five $\kappa$ values in the 10\% HR decay case as an example. 
The reduced $\chi_r^2$ is defined as $\chi^2$ divided by the number of degrees of freedom, 
$\nu = 52 \,(14 \times 4 - 4)$.  The $\chi_r^2$ curves exhibit a broad minimum around HTs of 25–40 MK. Models with very low or very high HTs are systematically disfavored, producing larger $\chi_r^2$  values. 
This indicates that the EUV observations constrain the HT to a relatively narrow range. 

Exceptionally, for the extreme $\kappa=2$, the $\chi_r^2$ goes to a minimum around the high HT of around 50~MK. This behavior can be understood in the context of the temperature sensitivity of the EUV passbands. 
The AIA channels primarily respond to plasma temperatures below $\sim$20~MK, 
and their sensitivity decreases rapidly at higher temperatures. 
As a result, the observations provide limited constraints on very hot components. 

In this temperature range, changes in $\kappa$ can partially compensate for the lack of temperature sensitivity by modifying the high-energy tail of the nonthermal electron distribution, which affects excitation and ionization rates. 
This behavior arises because the total energy of the electron distribution is determined by the temperature. For low $\kappa$ values, a larger fraction of the energy resides in the high-energy tail, which reduces the average energy of electrons in the quasi-thermal core. 

This explains why, in the extreme $\kappa = 2$ case, acceptable fits can occur at higher HTs around $\sim$50~MK. 
However, the preference for low $\kappa$ solutions in some regions of parameter space may also indicate that suprathermal electron populations are required to reproduce the observed emission. 
\citet{polito18} show that Fe~XXIV peaks at $\sim$18~MK assuming Maxwellian electron velocity distributions, but shifts to $\sim$40~MK with $\kappa=2$. This implies that the formation temperature of the same transition shifts to much higher temperatures under a low-$\kappa$ distribution, reflecting the enhanced high-energy tail of the electron population. It is consistent with our finding that the acceptable fits for $\kappa=2$ occur at higher $HT$.

Figure~\ref{fig:chi}(b) presents the minimum reduced $\chi_r^2$ as a function of $\kappa$ for each HR, 
where the minimum is taken over the HT grid. 
For a given HR, the $\chi_r^2$ variation with $\kappa$ is relatively shallow, 
with slightly lower $\chi_r^2$ values toward smaller $\kappa$. 
Overall, the dependence on $\kappa$ is weak, except in the lowest range ($\kappa=$ 2$–$4), 
where the trend becomes more noticeable.

This behavior indicates that a wide range of $\kappa$ values is allowed and that the AIA observations alone do not strongly constrain the electron distribution. However, models with lower $\kappa$ values ($\kappa=$ 2$–$4) show a more noticeable dependence. 
We further quantify this degeneracy using confidence regions in Section~\ref{subsec:accep}. 

\subsection{Ionization Evolution} 

To gain insight into the ionization behavior, 
we examine the time-dependent evolution of ion fractions for selected Fe charge states under NEI conditions with nonthermal electrons. 
The results shown here correspond to models within the acceptable range for the 10\% decay scenario, with $HR = 0.4$~erg~cm$^{-3}$~s$^{-1}$.

\begin{figure}
\centering
\includegraphics[width=50mm]{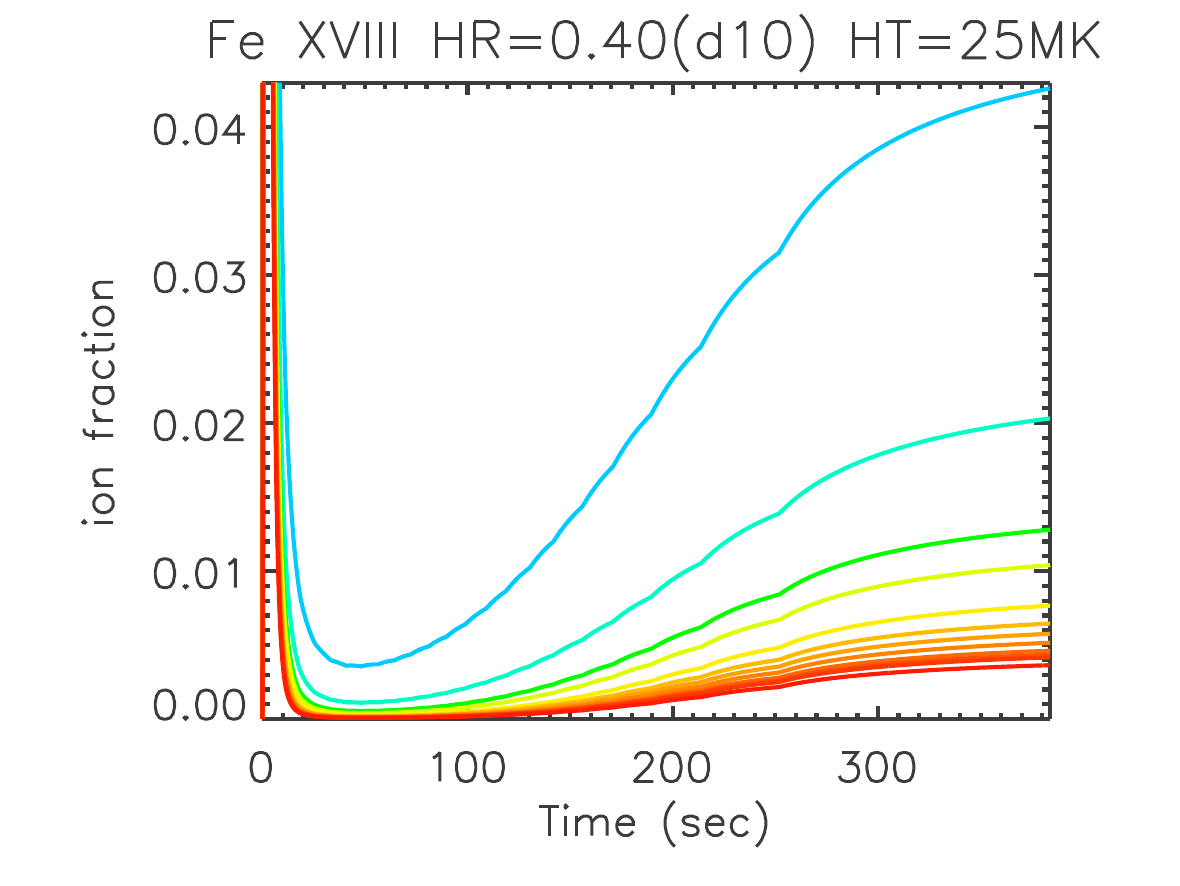}
\includegraphics[width=50mm]{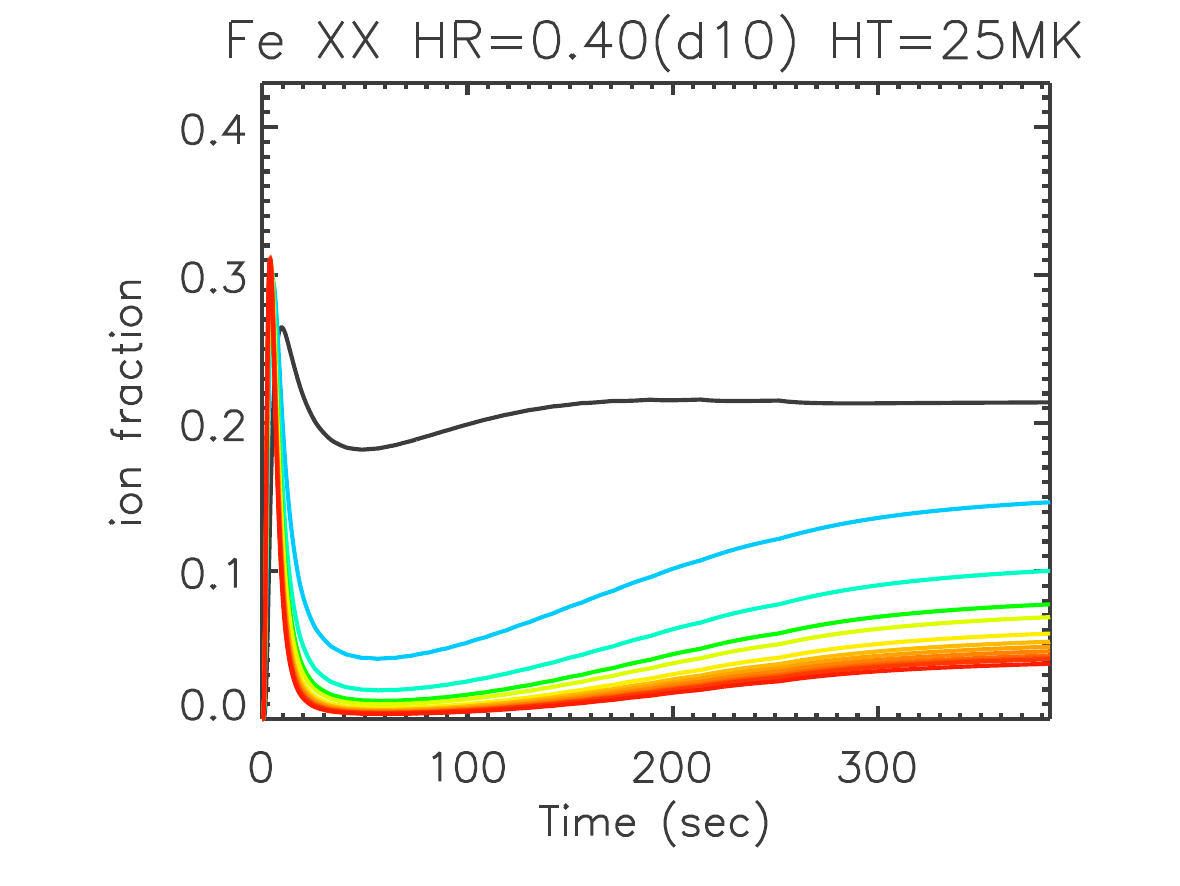} \\
\includegraphics[width=50mm]{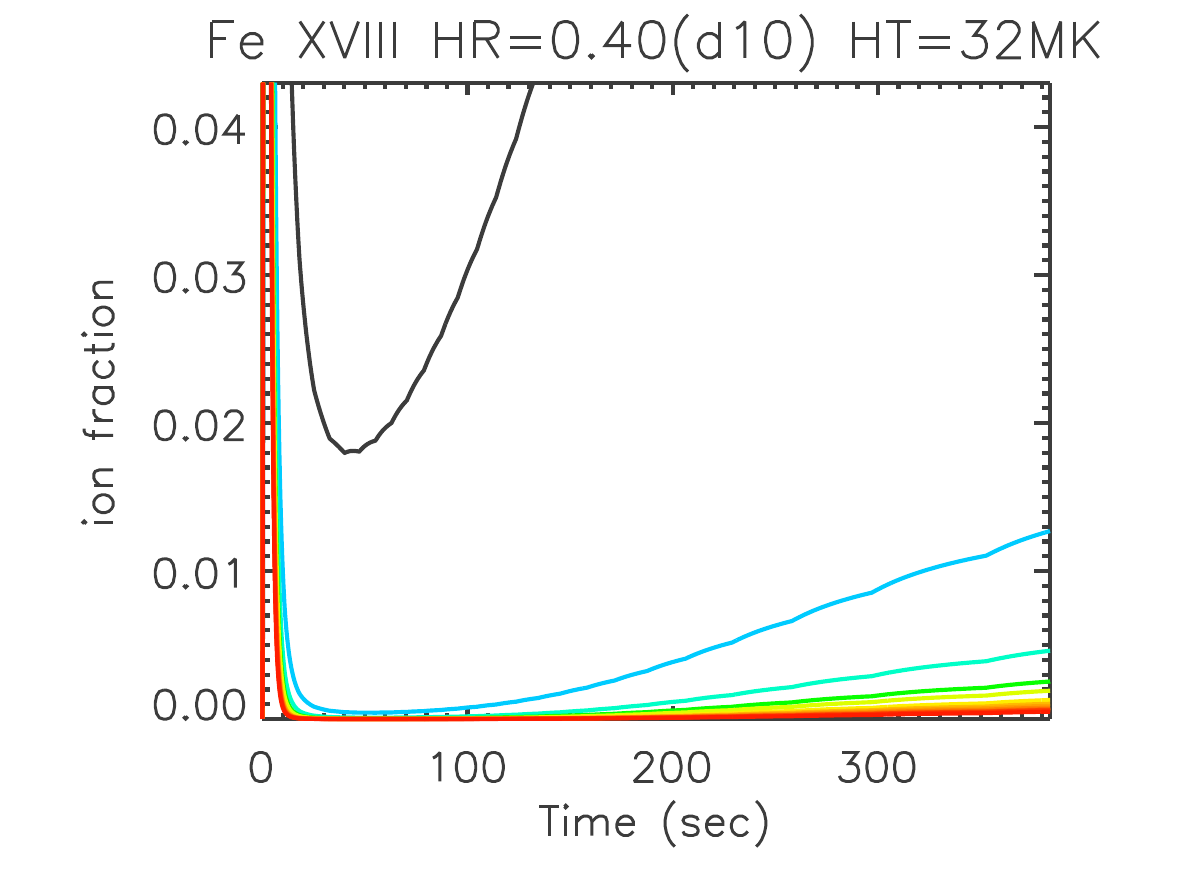}
\includegraphics[width=50mm]{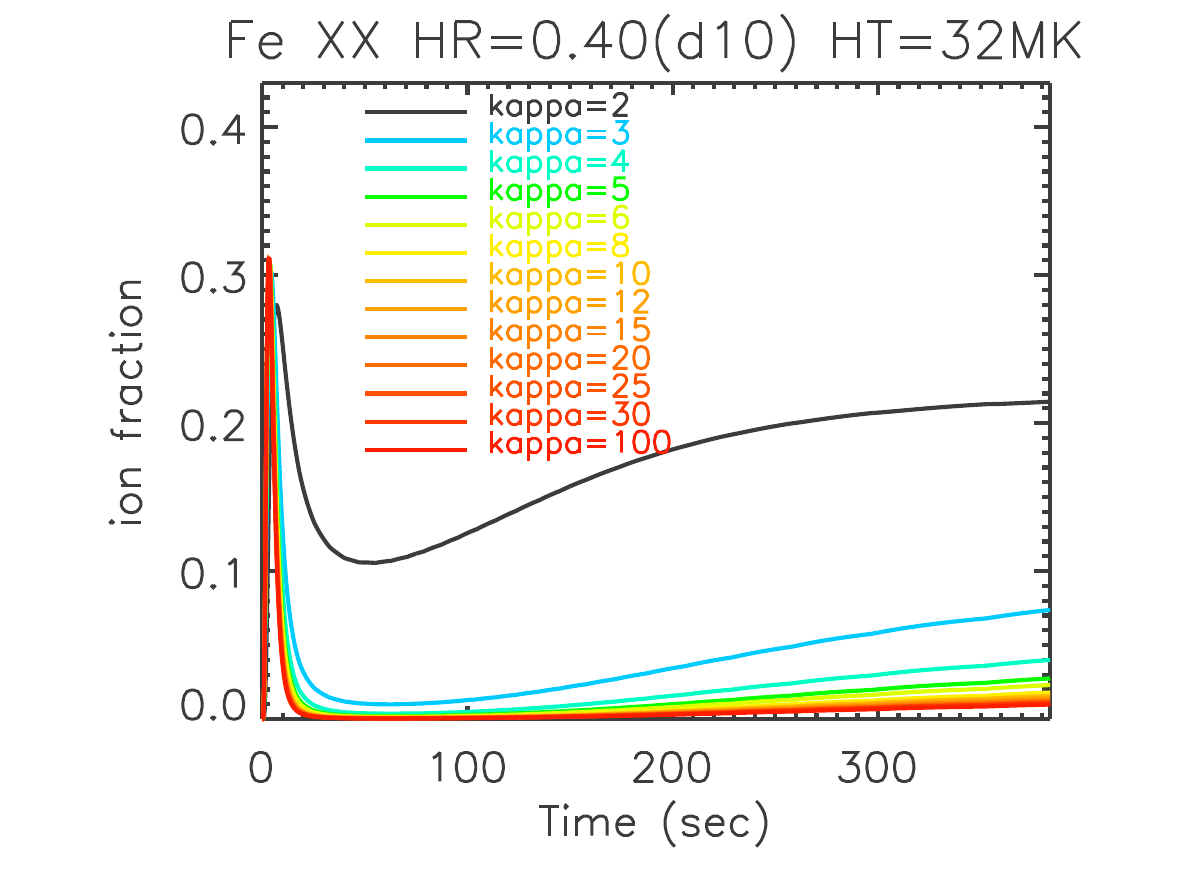} \\
\caption{Temporal evolution of Fe~XVIII and Fe~XX for different $\kappa$ values at a continuous heating rate of HR = 0.4 erg cm$^{-3}$ s$^{-1}$. The top and bottom panels correspond to HT = 25~MK and 32~MK, respectively. Colors denote different $\kappa$ values. Because the evolution becomes nearly indistinguishable for large $\kappa$ values, these cases are shown in similar red shades to highlight the differences among the low-$\kappa$ distributions.}
\label{fig:band94}
\end{figure}

Although Fe~XVIII, Fe~XX, and Fe~XXI are typically considered the dominant contributors to the 94 and 131~\AA\ channels during flares \citep{odwyer2010}, the Fe~XXIII ionization fraction is larger in most of our models. 
In Figure~\ref{fig:band94}, the Fe~XVIII fraction decreases as the plasma is ionized toward higher charge states. 
As a result, Fe~XX becomes relatively more significant in both the 25~MK and 32~MK cases. 
The overall fractions of both Fe~XVIII and Fe~XX are generally lower at $HT = 32$~MK than at 25~MK, 
indicating a more efficient redistribution toward higher charge states at higher temperature. 
For $\kappa = 2$, although the fractions are still lower at $HT = 32$~MK than at 25~MK, 
the charge states remain comparatively more persistent than in the higher-$\kappa$ cases.

\begin{figure}
\centering
\includegraphics[width=50mm]{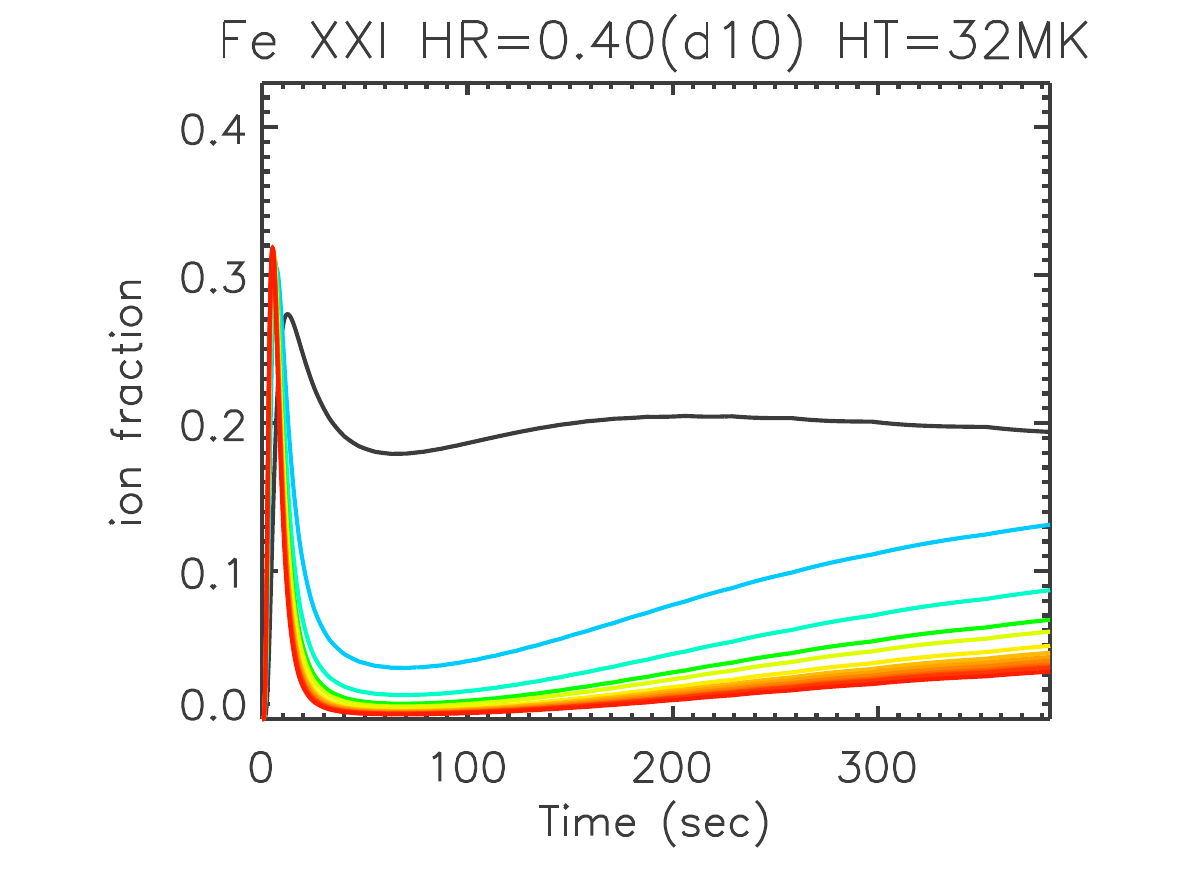}
\includegraphics[width=50mm]{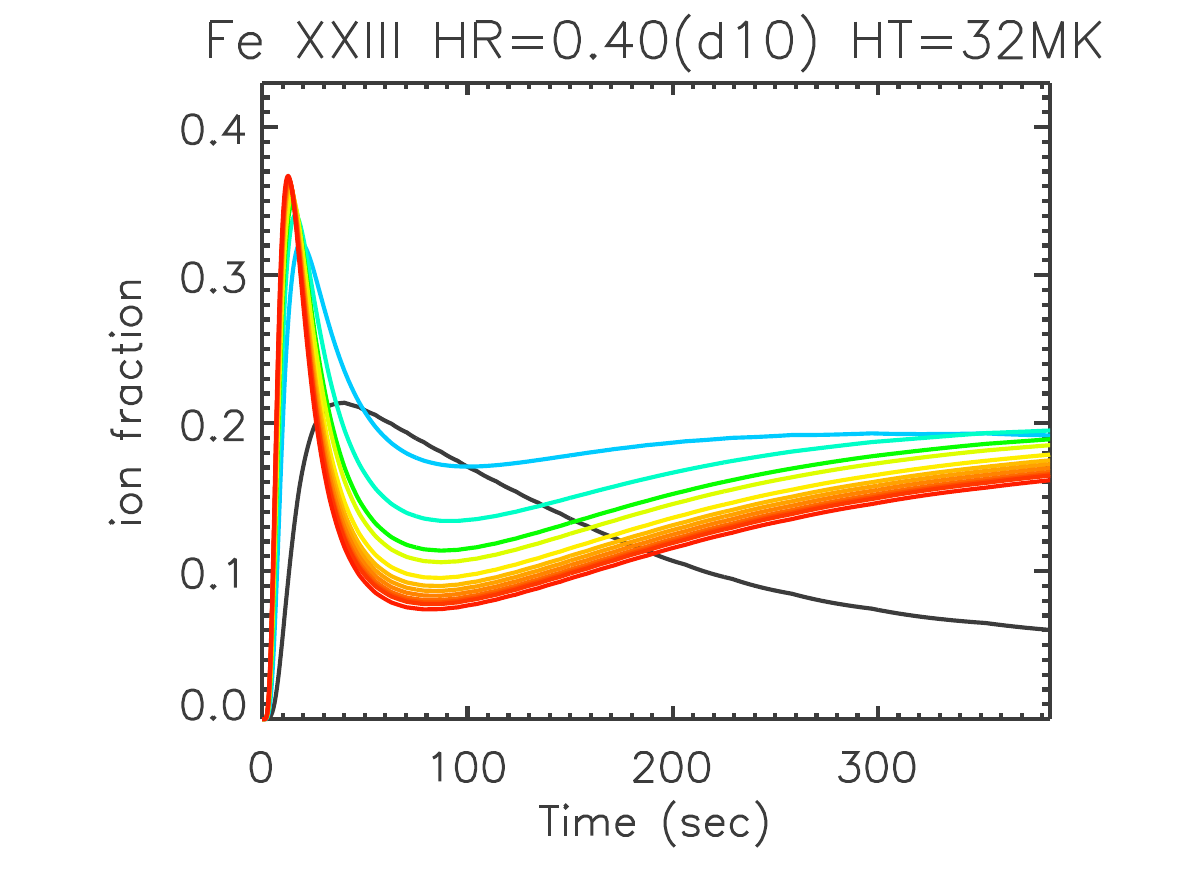}
\includegraphics[width=50mm]{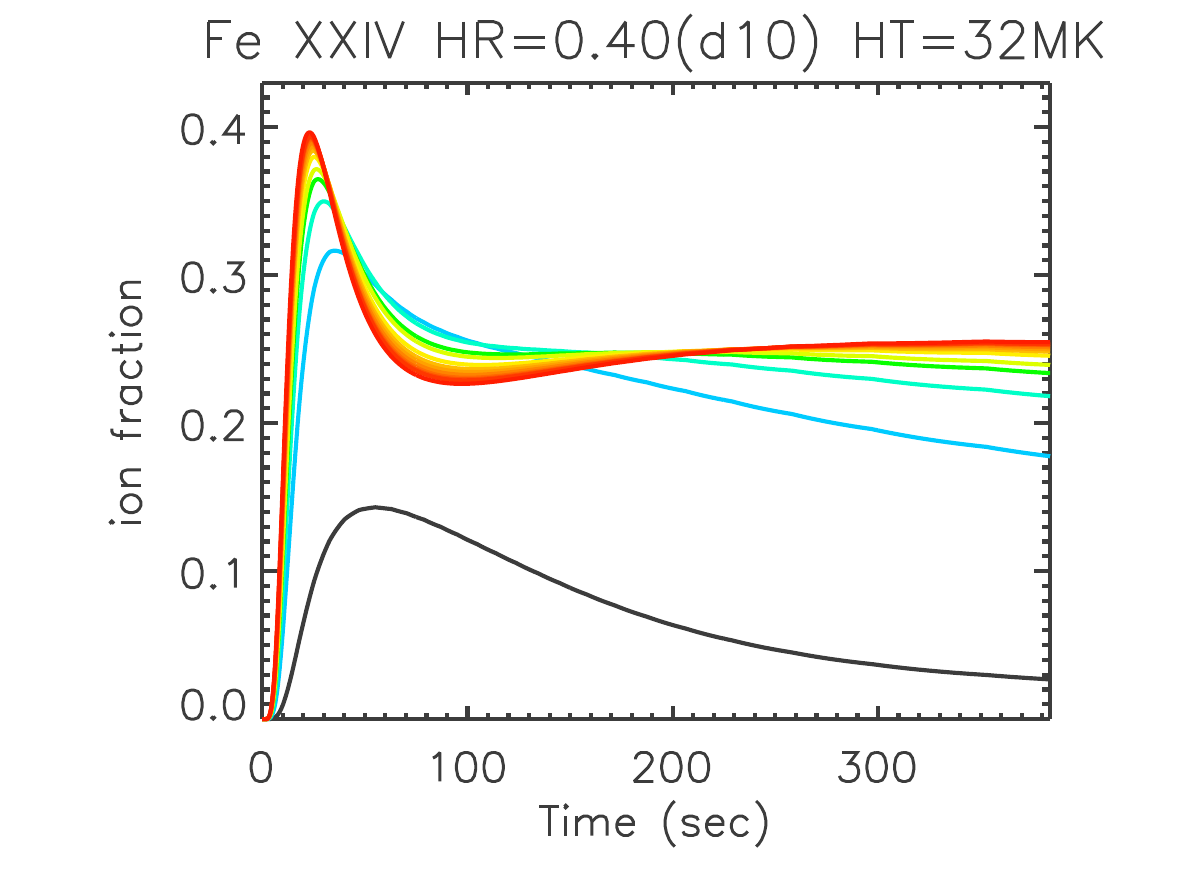} 
\includegraphics[width=50mm]{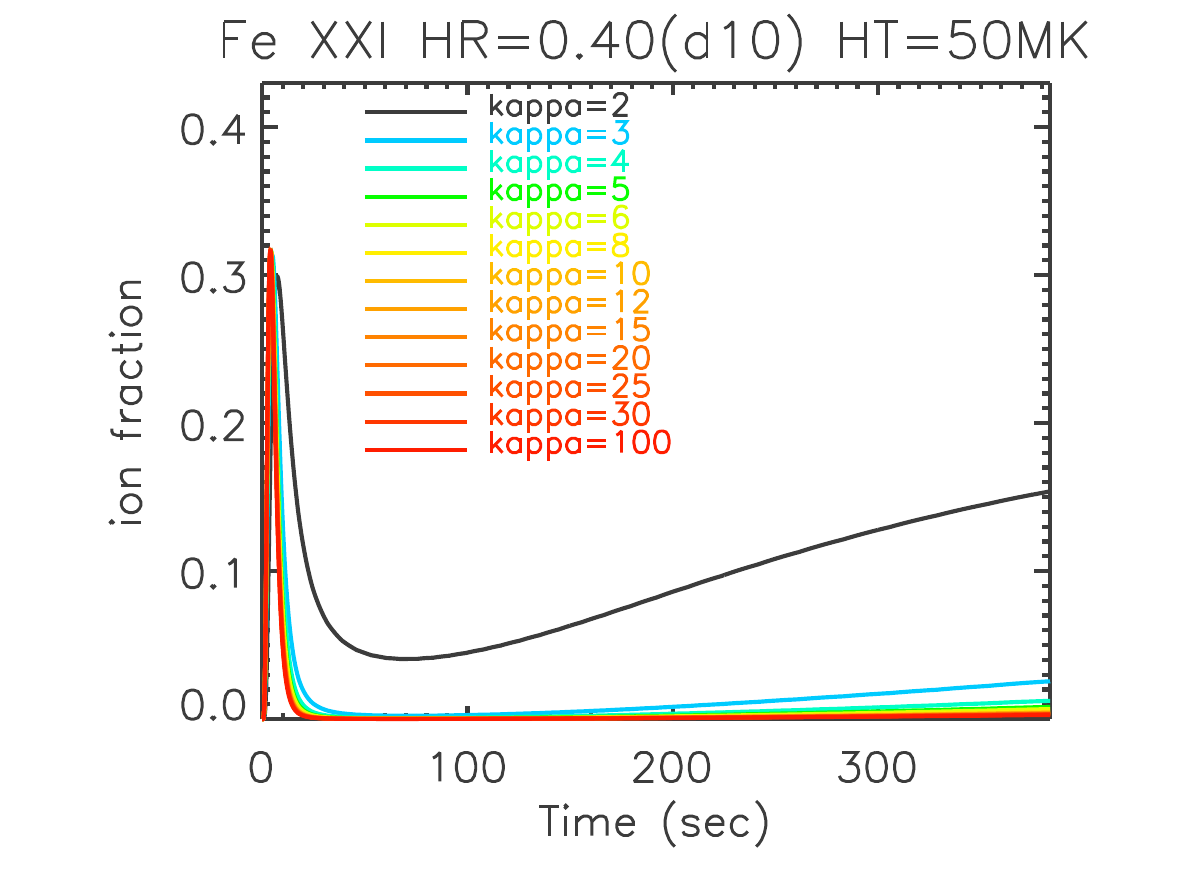}
\includegraphics[width=50mm]{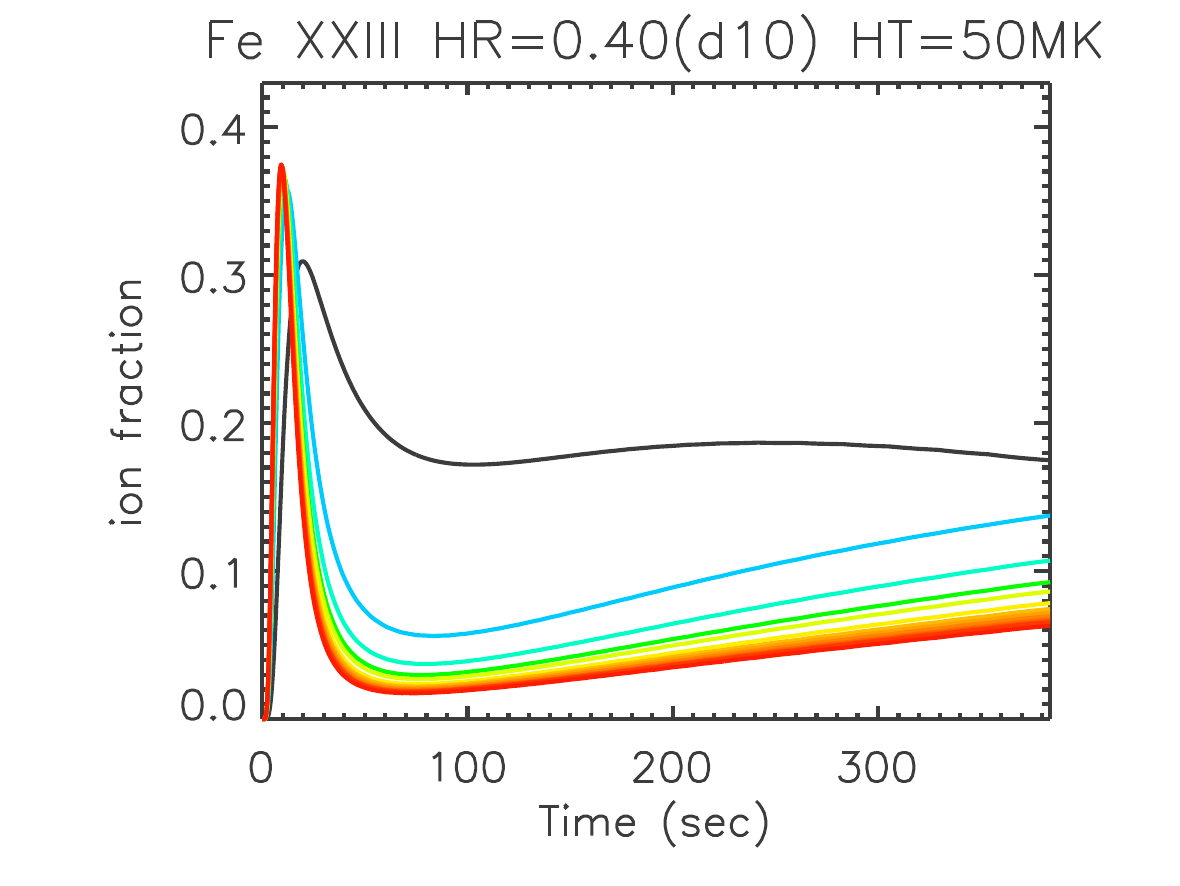}
\includegraphics[width=50mm]{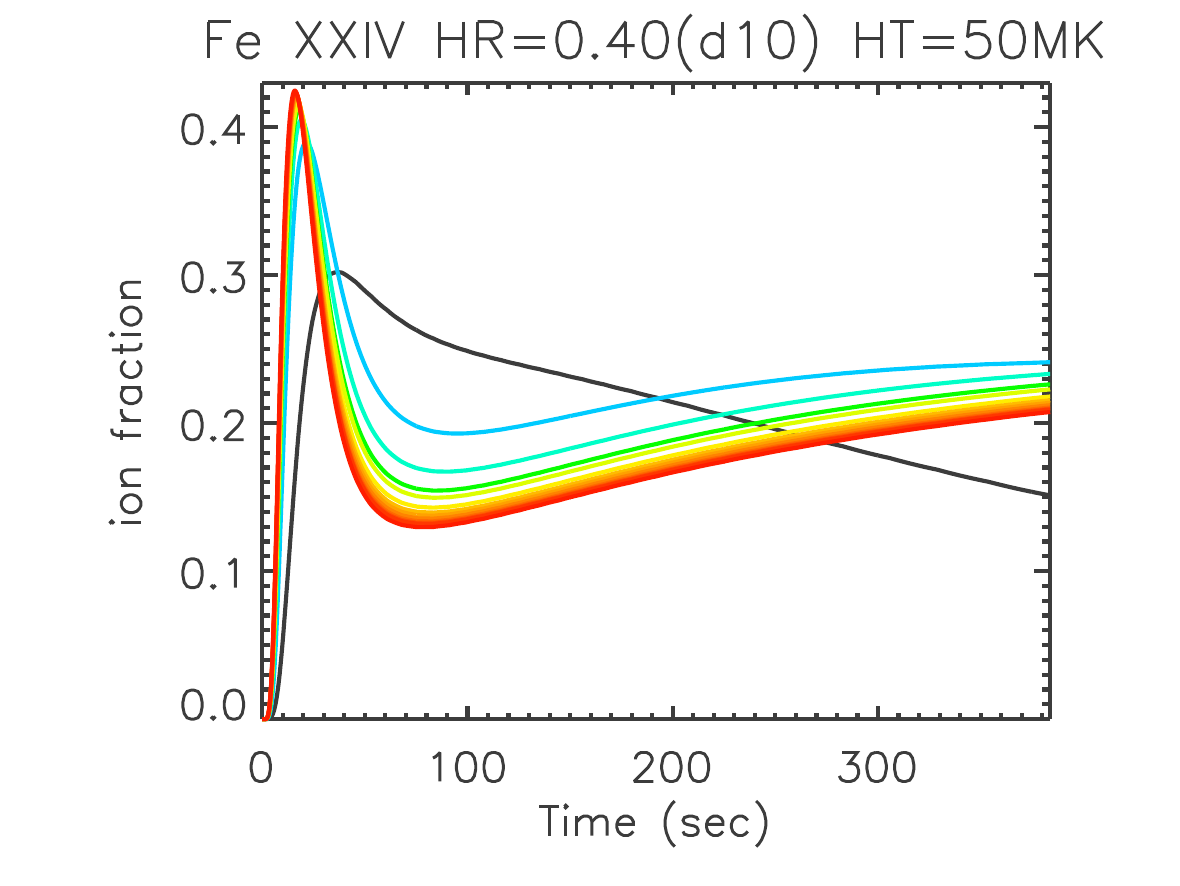}
\caption{Temporal evolution of Fe XXI, Fe XXIII, and Fe XXIV for different $\kappa$ values at HR = 0.4 erg cm$^{-3}$ s$^{-1}$. The top and bottom panels correspond to HT = 32~MK and 50~MK, respectively. The colors are the same as in Figure~\ref{fig:band94}.}
\label{fig:band131}
\end{figure}

In Figure~\ref{fig:band131}, the temporal evolution of Fe~XXI, Fe~XXIII, and Fe~XXIV shows a clear sequential progression toward higher charge states. Fe~XXI declines as ionization proceeds to higher stages, while Fe~XXIII and Fe~XXIV increase successively at later times, reflecting a time-delayed redistribution of the ion population under NEI conditions, where ionization and recombination cannot maintain equilibrium charge states. 

In both Figures~\ref{fig:band94} and \ref{fig:band131}, the ion fractions for lower $\kappa$ values evolve more gradually, 
indicating that the redistribution among charge states proceeds over a longer timescale compared to larger $\kappa$ cases. 
The ion fractions for $\kappa \gtrsim 4$–5 show broadly similar evolutionary patterns, 
with only minor differences in timing and amplitude, as also seen in Figure~\ref{fig:chi}(b). 
In contrast, for $\kappa = 2$ the progression toward higher charge states is noticeably delayed, 
suggesting that a stronger suprathermal tail may influence the ionization evolution in a qualitatively different manner. 

We emphasize that $\kappa$ is assumed to remain constant over the analyzed interval, 
and the inferred values should therefore be interpreted as effective, time-averaged parameters. 
Previous spectroscopic studies have reported evidence for strongly non-Maxwellian electron distributions in flare current-sheet environments, with $\kappa \lesssim 3$ inferred from line-profile analyses \citep{polito18}. 
The distinct behavior observed here for $\kappa = 2$ is qualitatively consistent with such interpretations, 
suggesting that suprathermal electron populations may play an important role in governing the charge-state evolution under NEI conditions.

\subsection{Statistically Acceptable Models}\label{subsec:accep}

To identify statistically acceptable models, we adopt the criterion 
$\Delta \chi^2 < 8.02$, corresponding to the 95.4\% confidence region 
for three free parameters ($HT$, $HR$, and $\kappa$), where $\Delta \chi^2 = \chi^2 - \chi^2_{\min}$. 
The confidence region is defined using the original $\chi^2$ values rather than the reduced $\chi_r^2$, 
since the $\Delta \chi^2$ threshold is based on differences in $\chi^2$ itself. 
Although the absolute value of $\chi^2$ increases with the number of data points, the $\Delta \chi^2$ threshold depends only on the number of free parameters. 
It therefore measures how far a model deviates from the best-fit solution in parameter space, 
rather than reflecting the total number of residual terms.

\begin{figure}
\centering
\includegraphics[width=50mm]{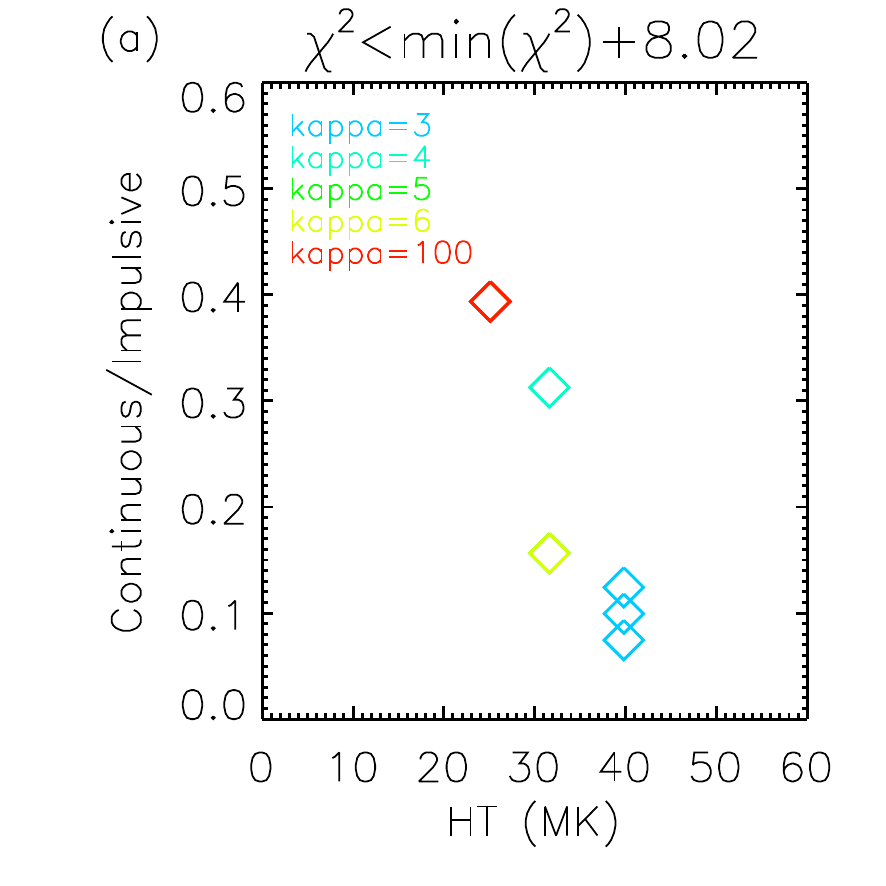}
\includegraphics[width=50mm]{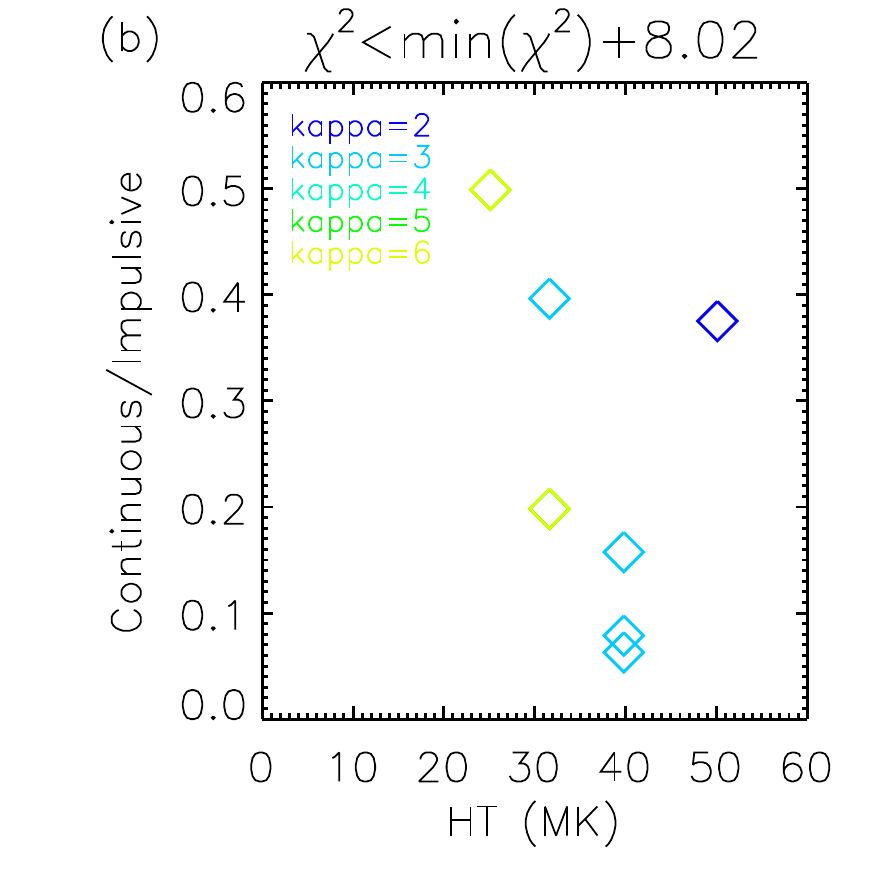}
\includegraphics[width=50mm]{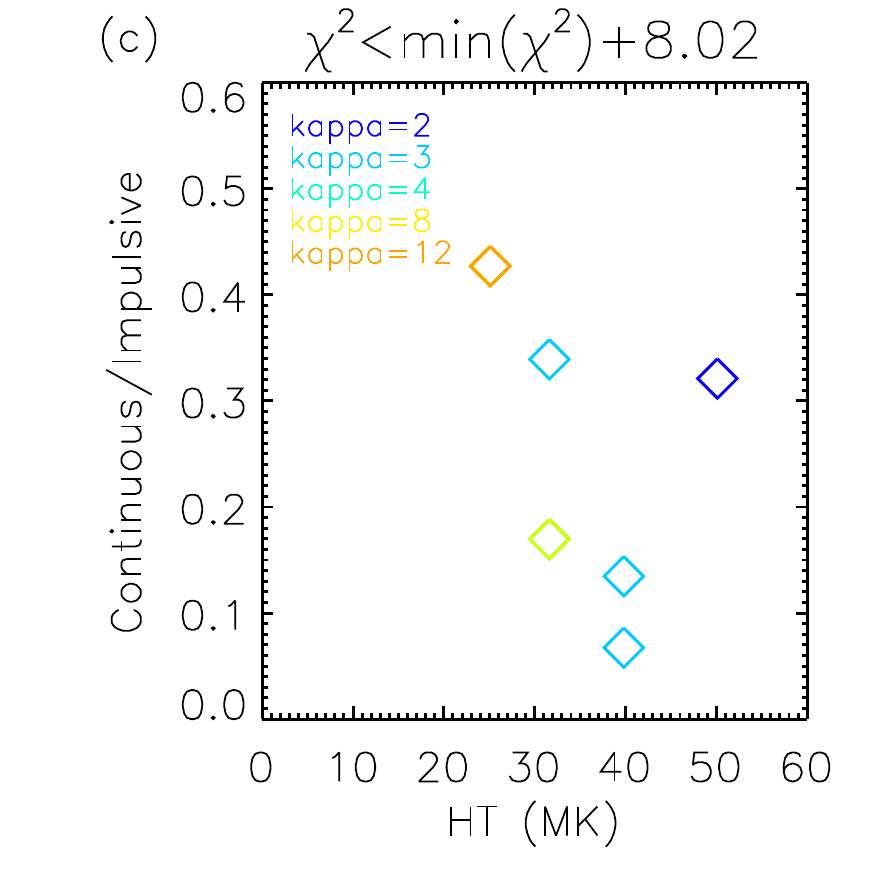}
\caption{
Ratio of total continuous to impulsive heating with $HT$ and $\kappa$ for models satisfying 
$\Delta \chi^2 < 8.02$,
where $\Delta \chi^2 = \chi^2 - \chi^2_{\min}$,
for three HR decay scenarios:
(a) 10\%, (b) 66\%, and (c) 80\%.
}
\label{fig:hratio}
\end{figure}

Figure~\ref{fig:hratio} shows the ratio of the total continuous heating energy integrated over the blob evolution to the impulsive heating energy as a function of $HT$ and $\kappa$ for models within the 95.4\% confidence region. 
Because multiple $\kappa$ values are allowed for the same $HT$ and heating ratio, the parameter ranges are listed  in Table~\ref{tab:rmodels}.
The heating ratio is evaluated using the impulsive heating energy per particle, $\frac{3}{2} k_B HT$, and the total continuous heating energy per particle, $\sum_{t=1}^{384} \frac{H_r(t)}{n(t)} dt$.

For the 10\% decay case, acceptable models cluster primarily at 
$HT = 25$–40~MK, spanning a broad range of $\kappa$ values at lower temperatures. 
When the total continuous-to-impulsive heating ratio is small, corresponding to strongly impulsive-dominated cases, 
only models with low $\kappa$ values ($\kappa \lesssim 3$) are statistically acceptable. 
In contrast, when the relative contribution of the total continuous heating increases, 
a broader range of $\kappa$ values becomes acceptable. 
This behavior suggests that strongly impulsive heating favors lower $\kappa$ solutions, indicating that a stronger suprathermal tail is required when the contribution of continuous heating is small.

In the 66\% and 80\% decay cases, the allowed region extends to 
$HT = 50$~MK only for very low $\kappa$ values ($\kappa = 2$). 
For larger $\kappa$, high temperatures are excluded from the confidence region. 
Although impulsive heating dominates in all acceptable models, temperatures near 50~MK are permitted only in the presence of a sufficiently strong suprathermal tail.

Overall, the confidence regions indicate that $\kappa$ is only weakly constrained by the observations. 
In particular, for the 10\% decay case, acceptable models span a wide range of $\kappa \sim 15$--100 at $HT = 25$~MK.  
This indicates that the electron distributions close to Maxwellian are consistent with the observations. 
High temperatures are only allowed for low $\kappa = 2$--3, with $HT \sim 50$~MK, as the enhanced high-energy tail in low $\kappa$ distributions is required to match the observations. 
In addition, impulsive heating dominates over continuous heating in all statistically acceptable models.

\begin{figure}
\centering
\includegraphics[width=50mm]{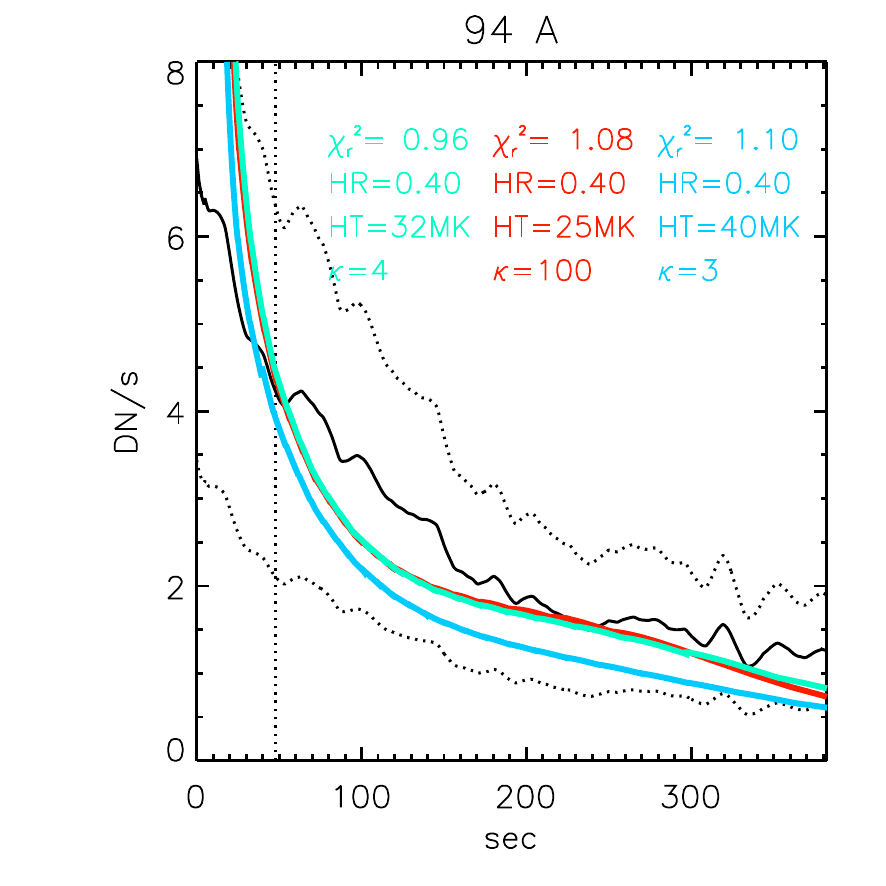}
\includegraphics[width=50mm]{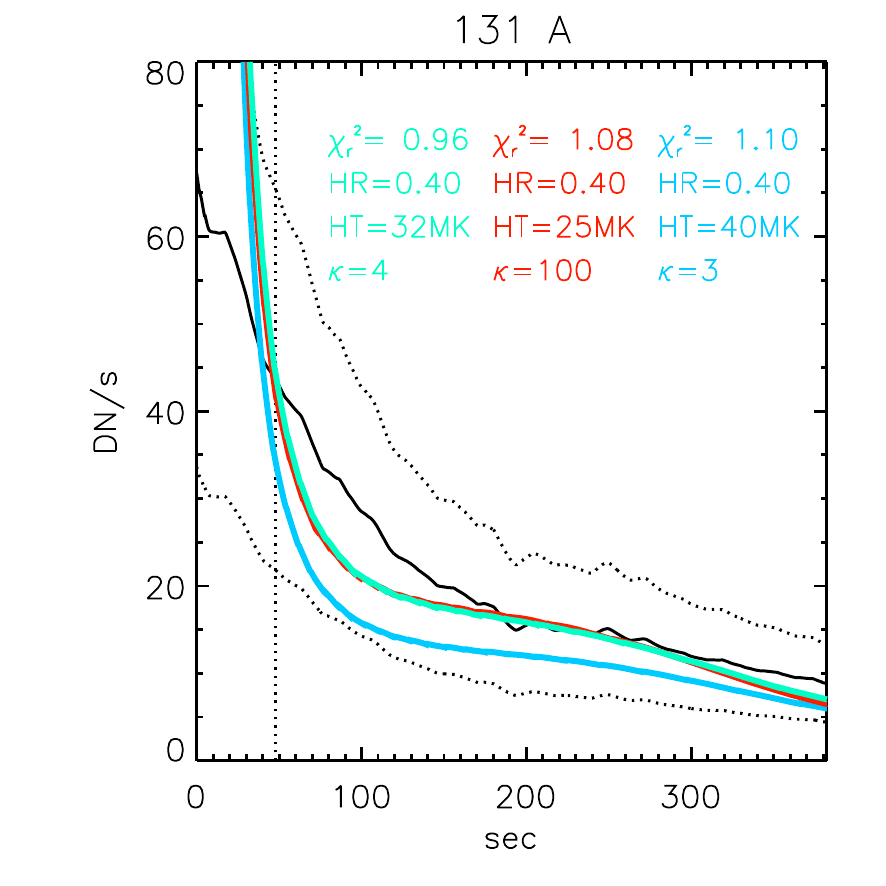} \\
\includegraphics[width=50mm]{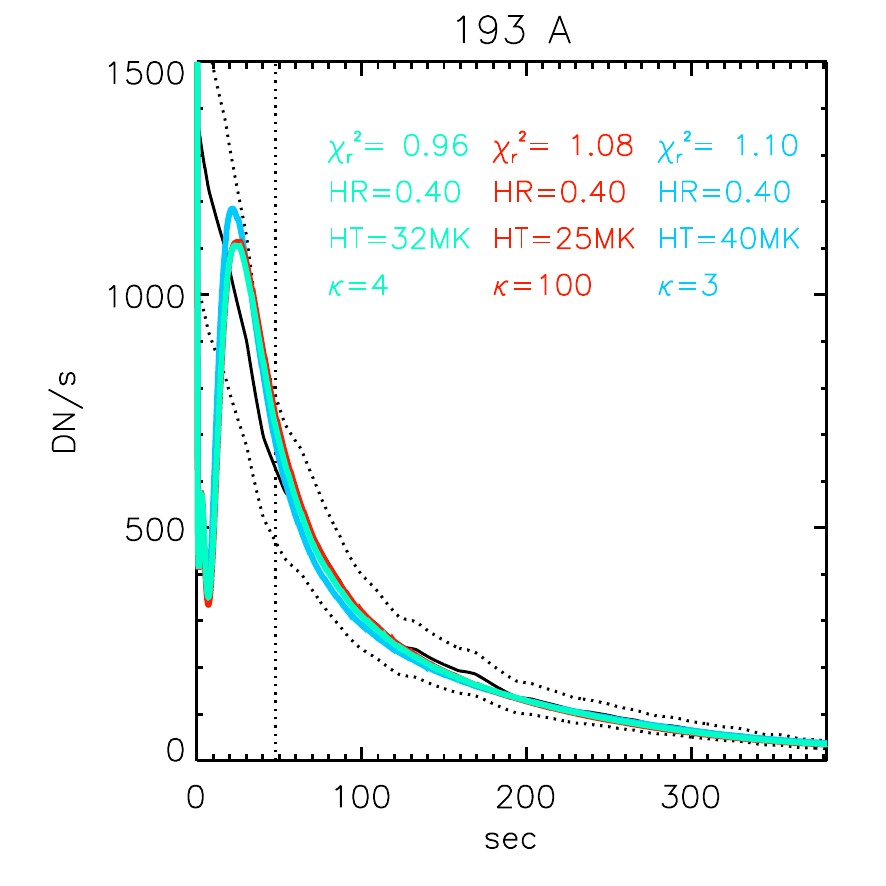}
\includegraphics[width=50mm]{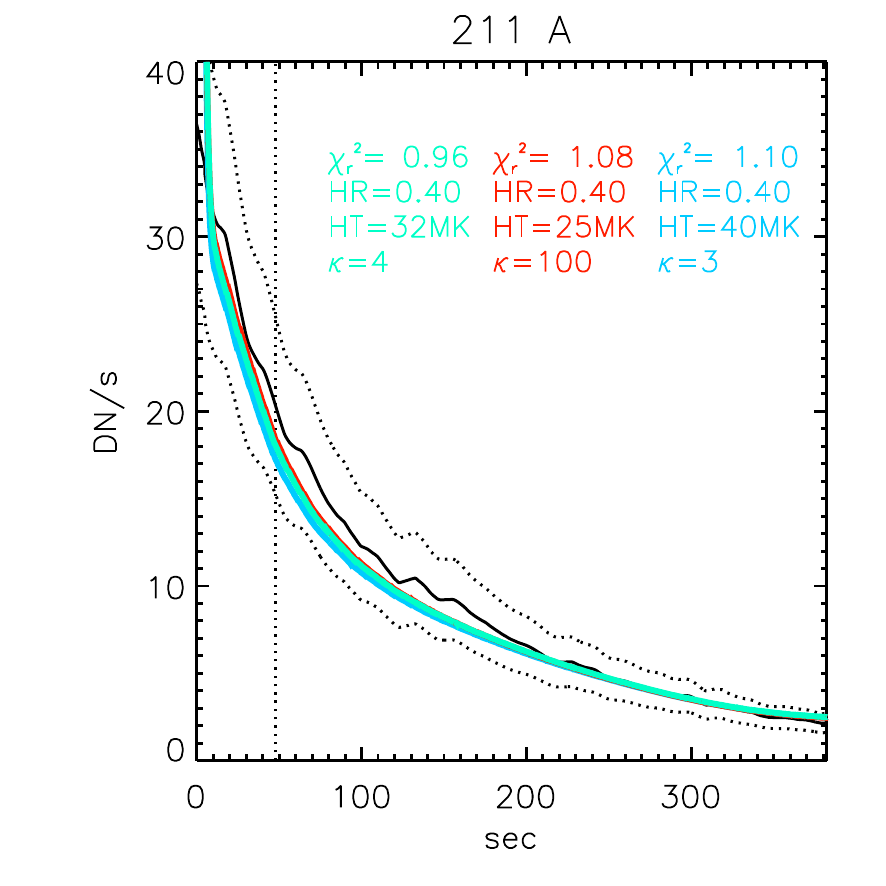}
\caption{Comparison of the observed light curves and the representative models for the 10\% decay case. 
Black solid curves denote the observations, dotted curves their uncertainties, and red solid curves the representative models multiplied by the band normalization factor $a_{\mathrm{band}}$ (Eq~\ref{eq:residual_vector} and Table~\ref{tab:bmodels}). 
Vertical dotted lines mark the starting time, 48 s after the first data point, used in the $\chi_r^2$ calculation. 
}
\label{fig:bestmodels}
\end{figure}

Figure~\ref{fig:bestmodels} compares the observed light curves with three representative models for the 10\% decay case, 
which are within the statistically acceptable range.
The models reproduce the subsequent decay phase within the observational uncertainties across all four EUV passbands. Comparable agreement is obtained for the other decay scenarios.
The parameters for the representative models for the three HR decay scenarios are summarized in Table~\ref{tab:bmodels}.
We also show the band normalization factors $a_{band}$. 
The normalization factors span 0.5-1.4, with the 131, 193, and 211~\AA\ channels remaining close to unity. 
The systematically larger scaling in the 94~\AA\ channel ($\sim$2.5–3.1) 
may reflect residual uncertainties in its high-temperature response 
or atomic emissivities rather than deficiencies in the overall model structure. 
At these high temperatures, Fe XVIII contributes negligibly to the 94 Å band, while Fe XX dominates. Because the principal Fe XX contribution is an intercombination doublet with relatively large atomic uncertainties, this may account for the enhanced scaling required for the 94 Å channel. Since the temperature constraints are primarily derived from the relative temporal evolution across multiple passbands, this does not significantly affect the inferred temperatures. The individual line contributions are shown in Figure~\ref{fig:band94}.

In addition, differences in density sensitivity among the EUV passbands under NEI conditions 
may contribute to the required scaling. 
The band-dependent normalization factors therefore likely compensate for a combination of 
instrumental response uncertainties and plasma-condition differences 
not fully captured by the simplified model assumptions. 

\begin{figure}
\centering
\includegraphics[width=50mm]{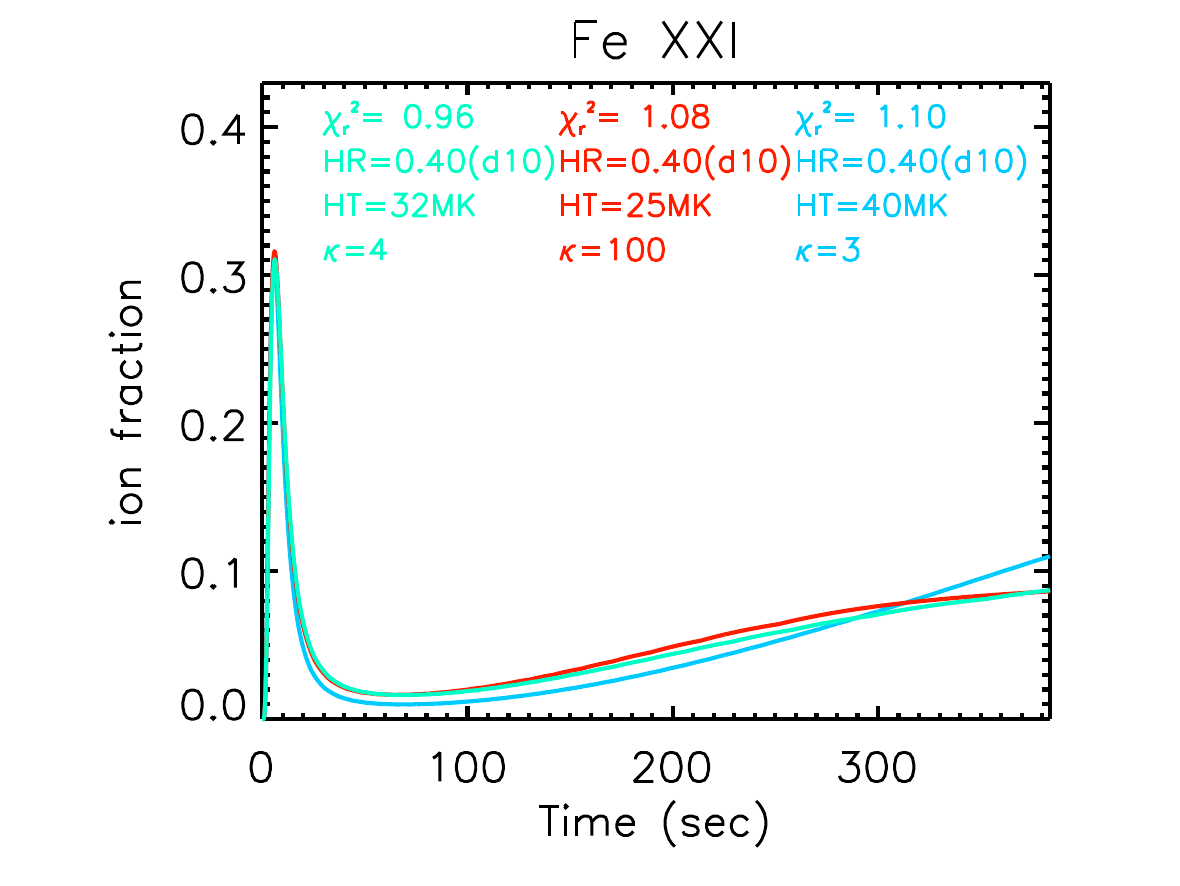}
\includegraphics[width=50mm]{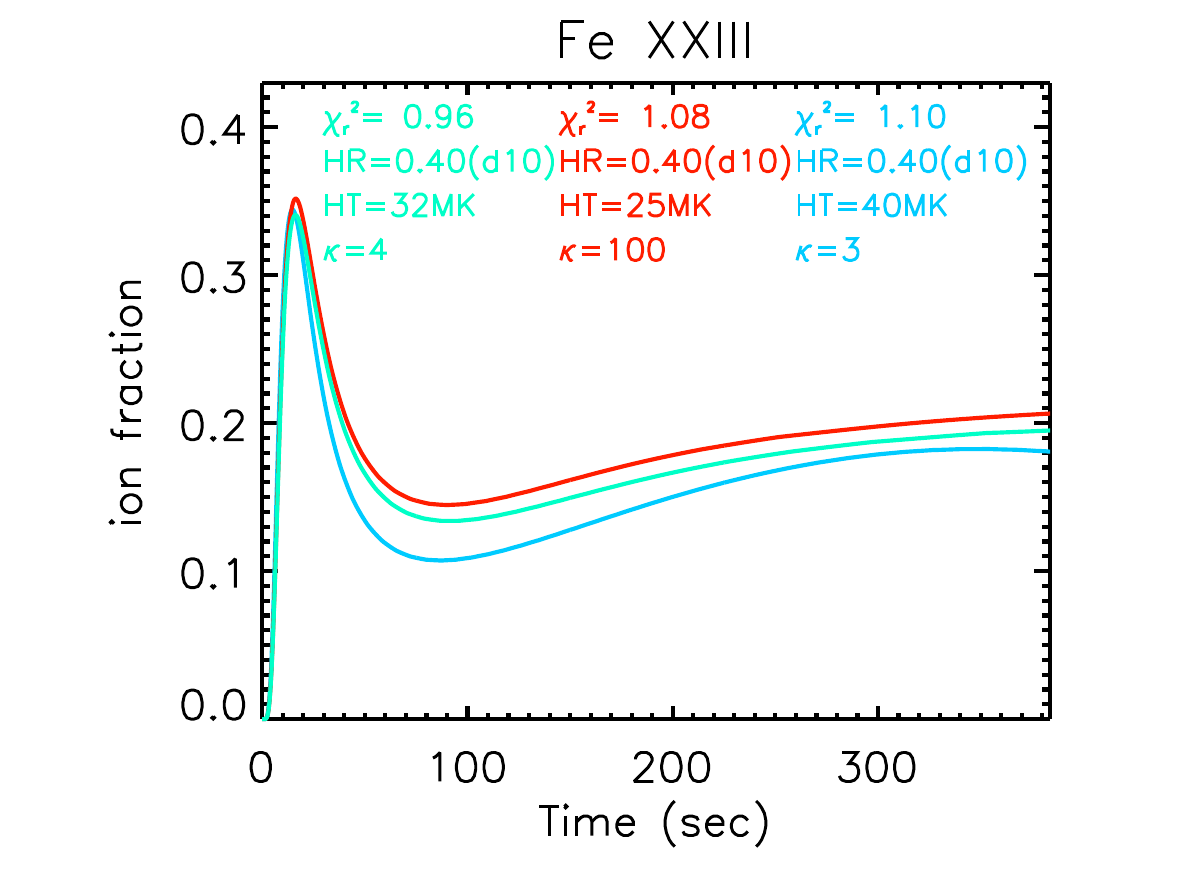}
\includegraphics[width=50mm]{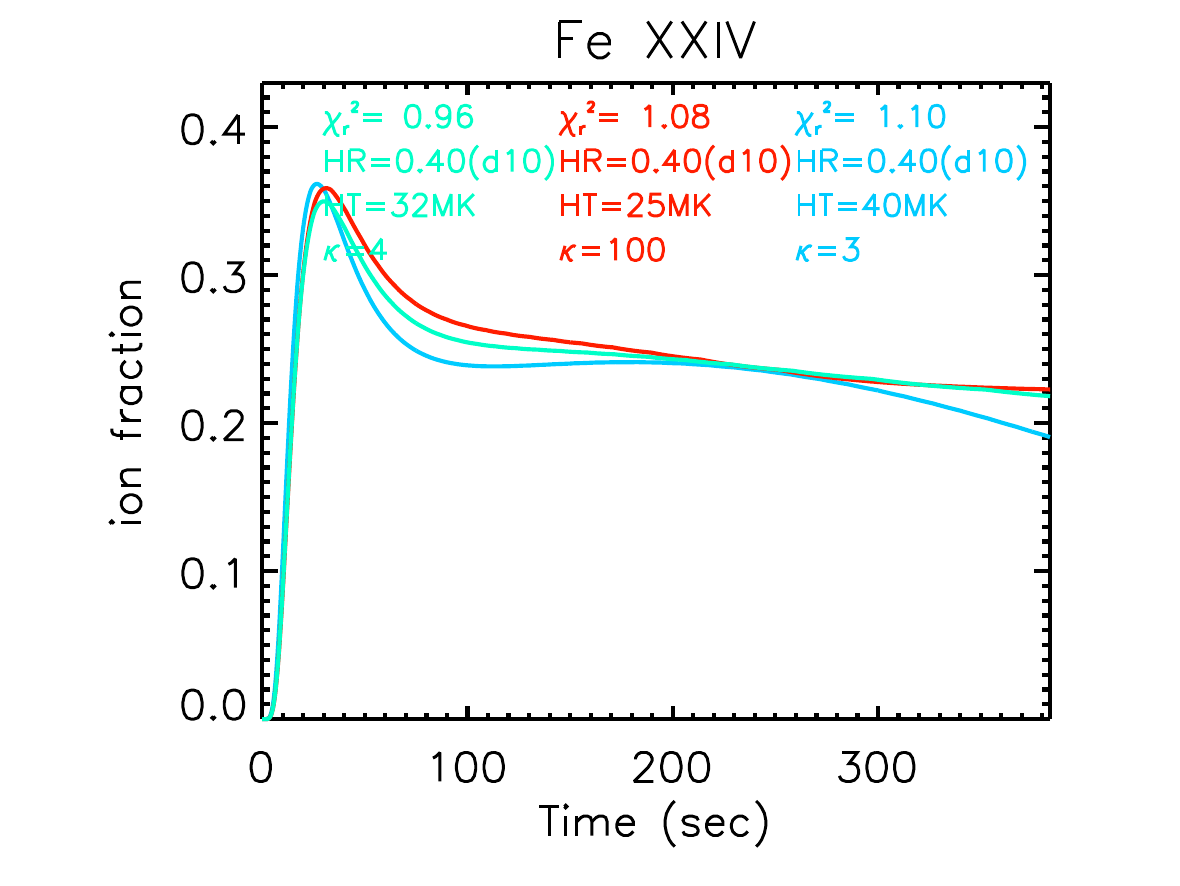} 
\includegraphics[width=50mm]{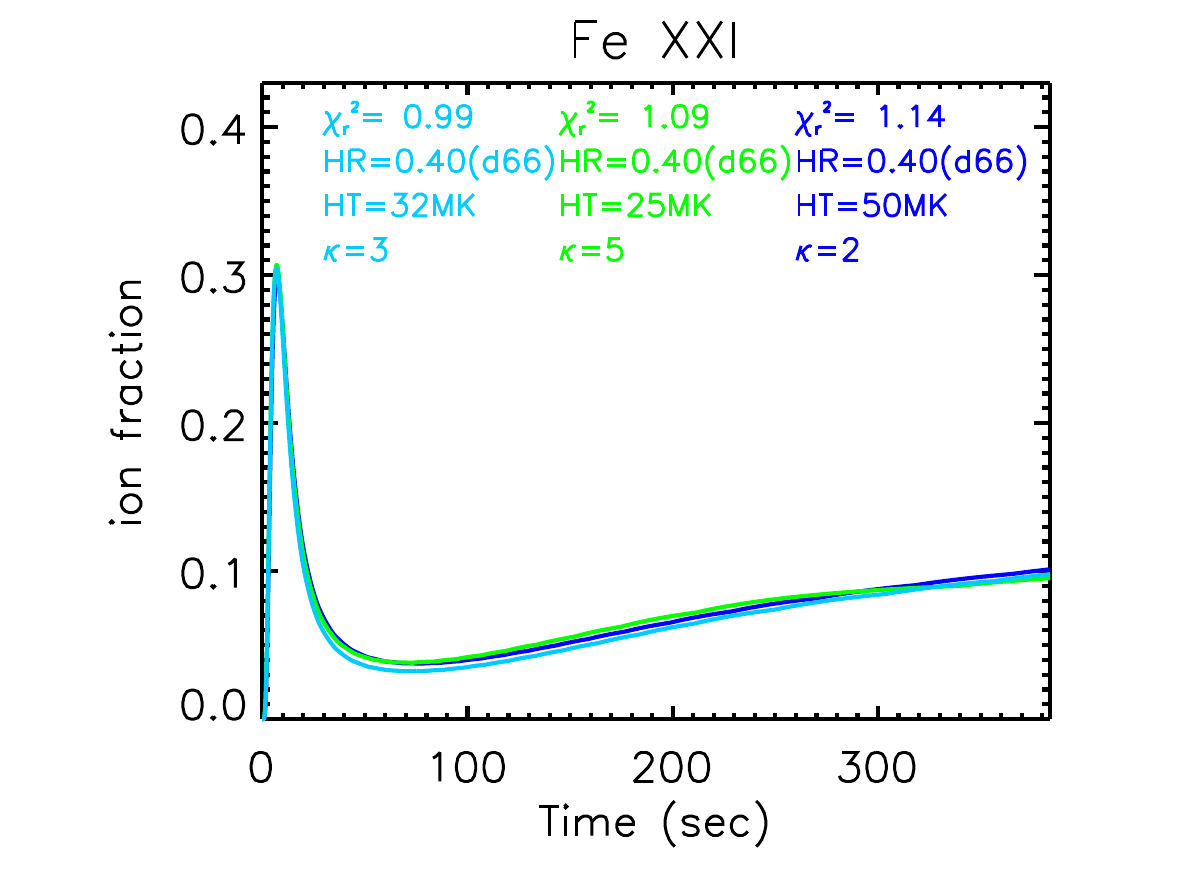}
\includegraphics[width=50mm]{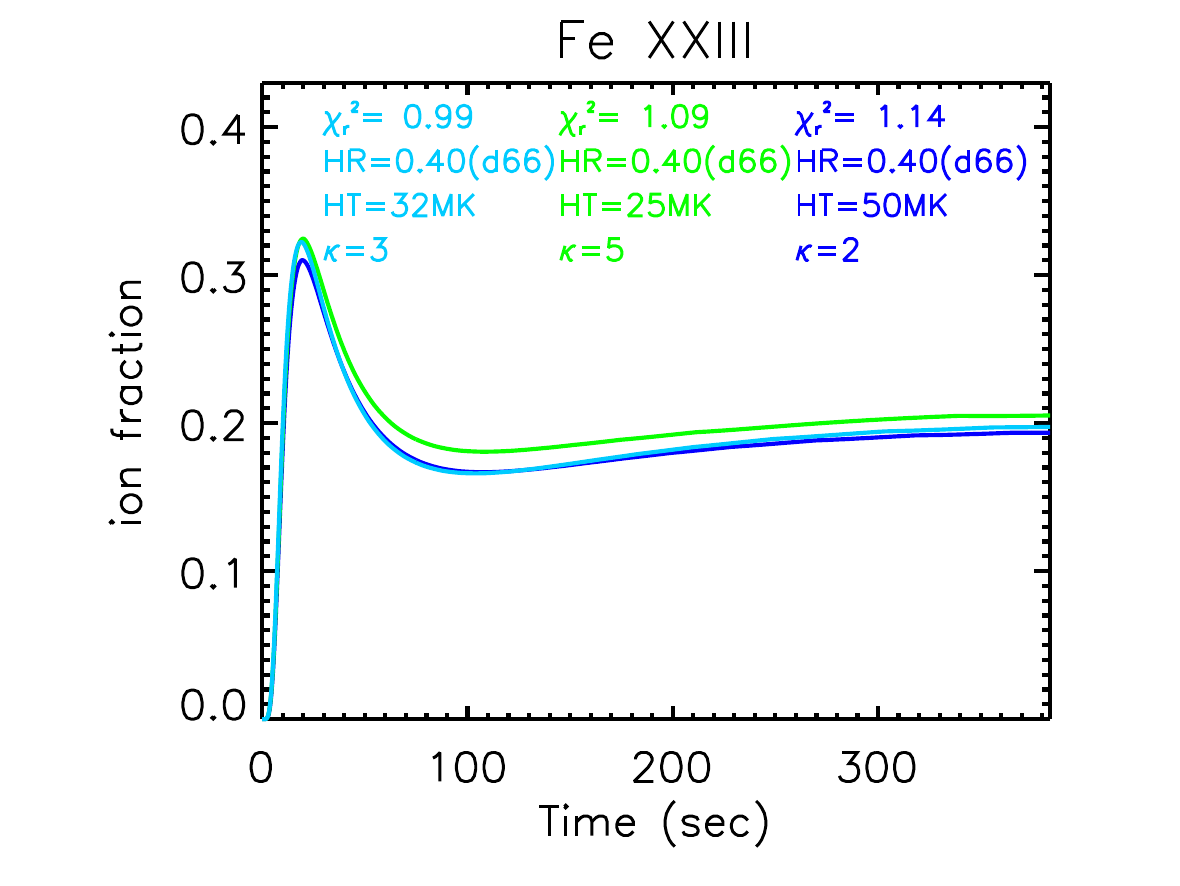}
\includegraphics[width=50mm]{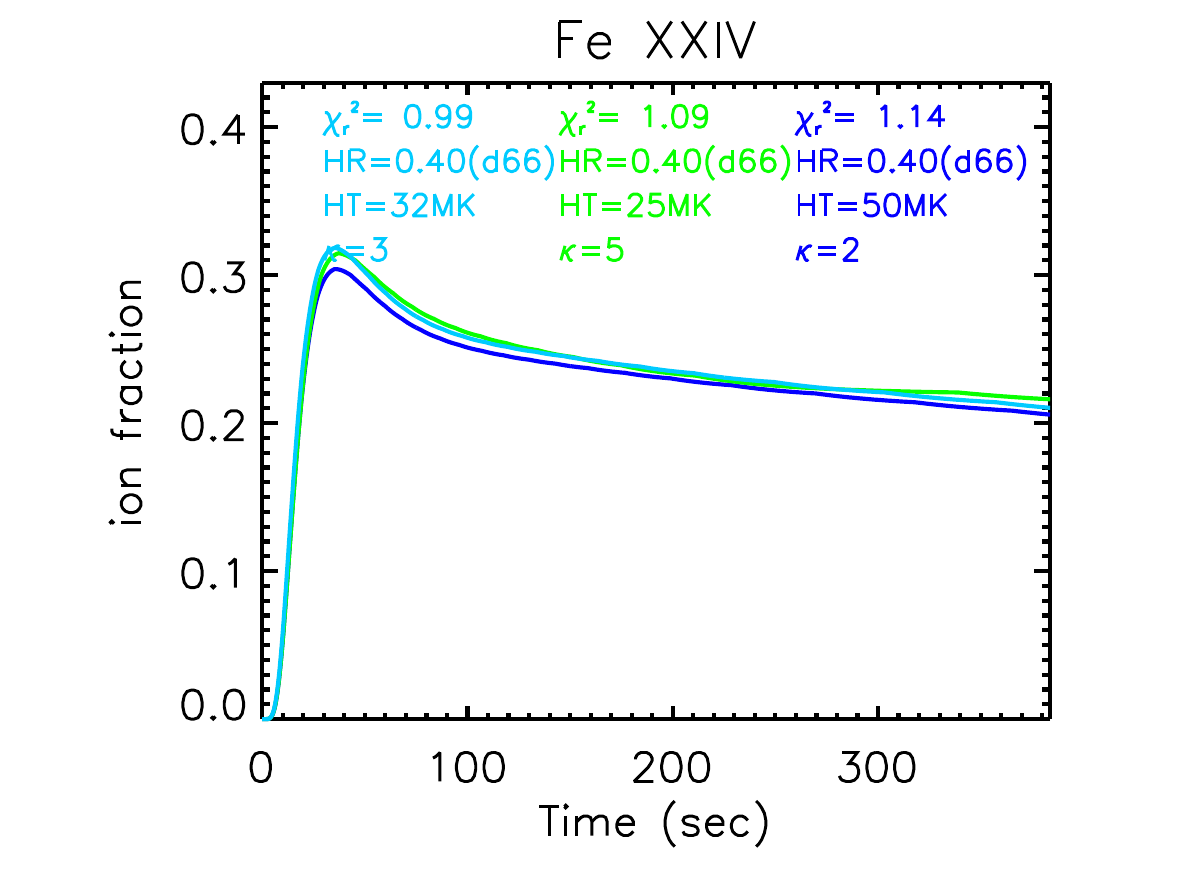}
\includegraphics[width=50mm]{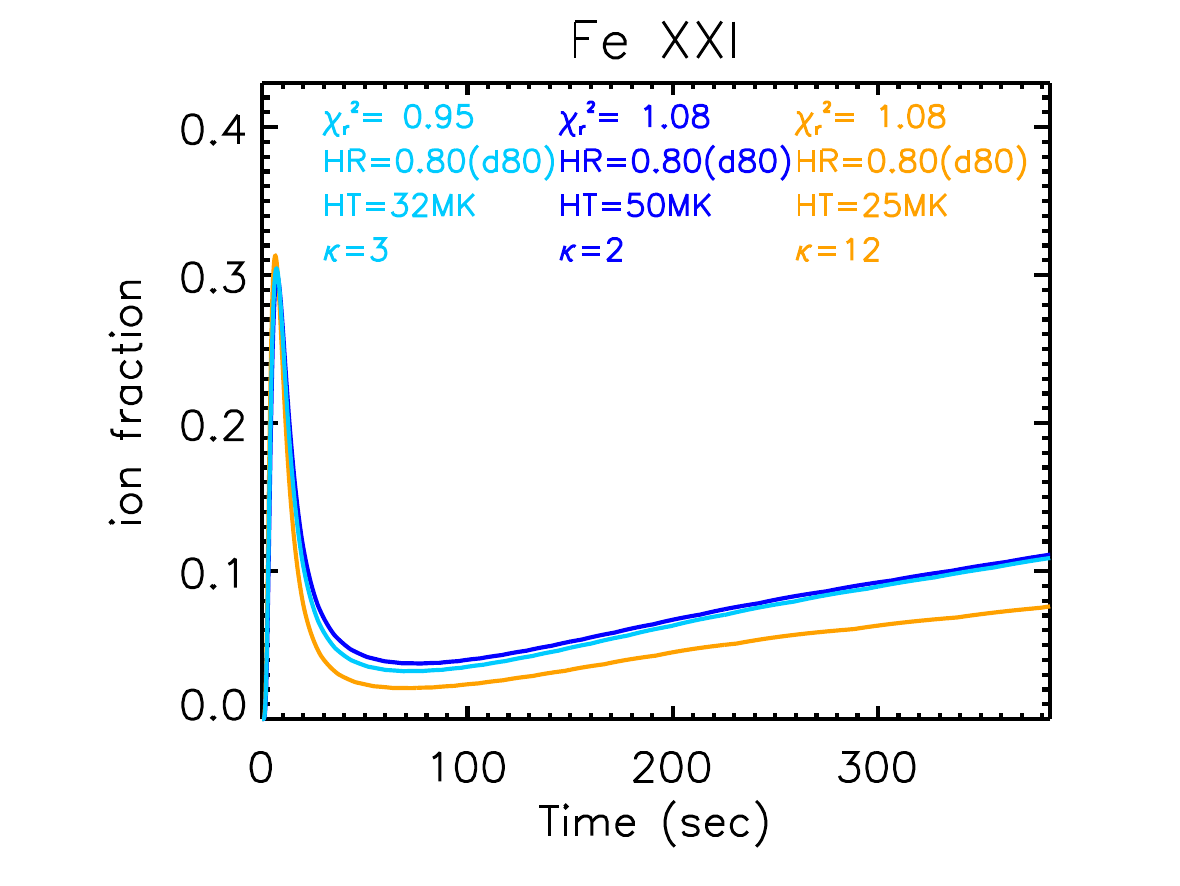}
\includegraphics[width=50mm]{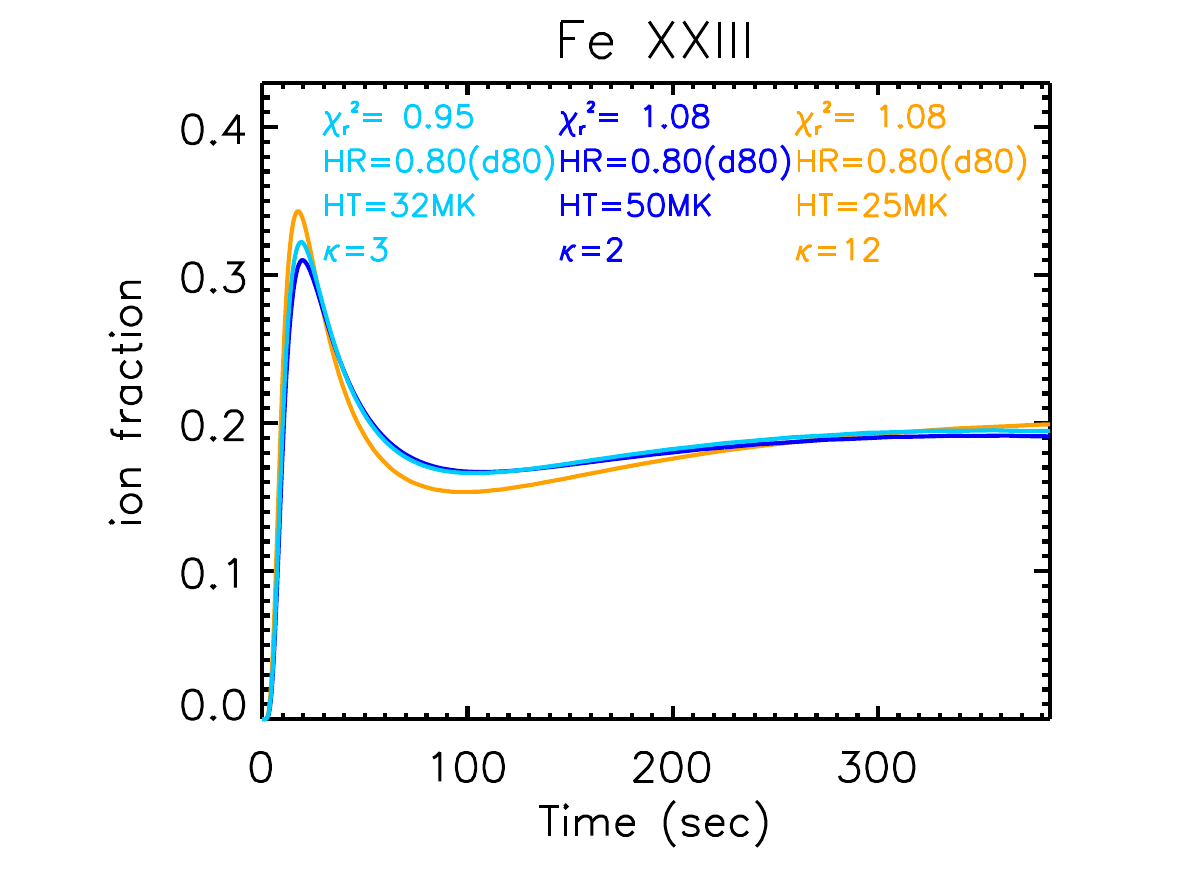}
\includegraphics[width=50mm]{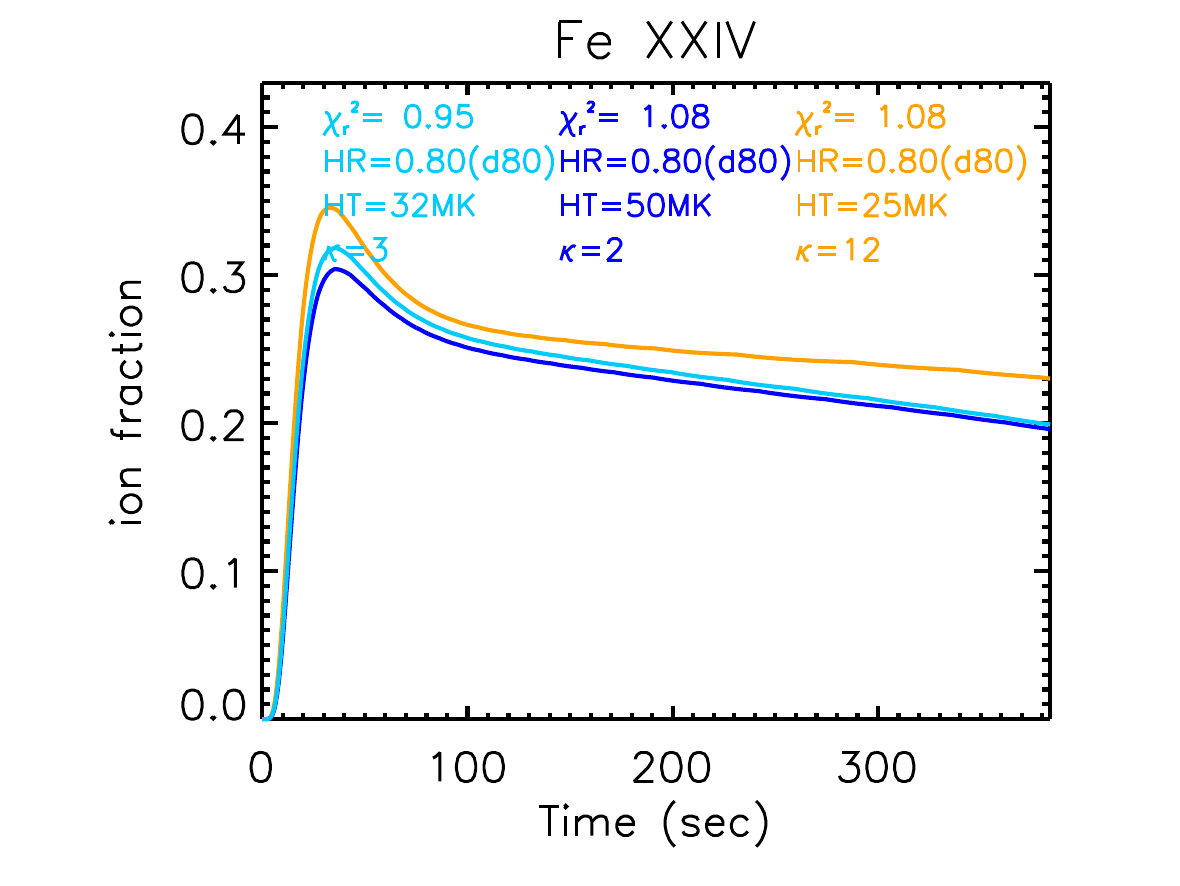}
\caption{Temporal evolution of Fe XXI, Fe XXIII, and Fe XXIV for three representative models in each decay scenario. 
The top, middle, and bottom panels show the evolution for 10\%, 66\%, and 80\% decay scenarios, respectively.  
}
\label{fig:model_ions}
\end{figure}

To understand why the models with lower $\kappa$ and high temperatures can reproduce the observations, we investigate in details the ion fraction evolution of three representative models for each decay scenario. 
Figure \ref{fig:model_ions} shows the Fe XXI, Fe XXIII, and Fe XXIV evolution for the three models in each decay scenario. 
The temporal evolution of the ion fractions shows that the overall profiles, including the rise, peak, and decay phases, are broadly similar among the different models. In particular, models with lower $\kappa$ and higher temperatures show profiles similar to those with higher $\kappa$ and lower temperatures. 

At a fixed temperature, ion fractions with lower $\kappa$ evolve more slowly than those with higher $\kappa$. However, at higher temperatures, the ion fractions evolve more rapidly, as seen in Figure~\ref{fig:band131}. Thus, the slower evolution in low $\kappa$ models can be compensated by higher temperatures, leading to similar ion fraction evolution. This behavior explains why lower $\kappa$ models can reproduce the observed ionization evolution even at relatively high temperatures. 

\section{Discussion}

\subsection{Cooling processes}

Thermal conduction and radiative cooling could contribute to the energy balance in the plasma sheet. 
\citet{Kittrell25} estimated a thermal conduction cooling rate comparable
to the heating rates inferred in this study. The temperature of the blob does not stand out compared to
the regions above and below it, and it is not obvious whether conduction would heat or 
cool the blob.  However, if the blob is a plasmoid, it is magnetically isolated from the 
surrounding plasma,  cutting off any thermal conduction.  For radiative cooling,  
we compare the 384-second flow time with the radiative cooling time scale 
$\tau_{\rm rad} \sim 5 k_{B}T/(n_e \Lambda(T)$), where $\Lambda(T)$ is the optically thin radiative loss function.  
For a density of 10$^{10}$ cm$^{-3}$ and a temperature of 25 MK, the $\tau_{\rm rad} \sim 5000$ s, which is much longer than the flow time scale. 
This is consistent with previous estimates \citet{Kittrell25} and supports the assumption that 
radiative losses do not significantly affect the temperature evolution in our model.   

\subsection{Heating during magnetic reconnection}

As described in Section~\ref{subsec:temperature}, the parameterized continuous heating term in this study 
represents the combined effects of multiple physical processes, including wave dissipation, magnetic reconnection, turbulent heating, conductive flux divergence, and shocks associated with reconnection outflows. 
A direct comparison between the heating rates and predictions from theoretical magnetic reconnection models is not straightforward. 

Nevertheless, the continuous heating term can be interpreted as energy release processes in the reconnecting current sheet, 
such as turbulence, plasmoid interactions, or wave dissipation. 
Considering a turbulent velocity up to 150~km s$^{-1}$ at low heights in \citet{warren18}, 
that corresponds to about 70~eV per particle, or about 0.8 MK. 
The dissipation time for hydrodynamic turbulence is $t_{\rm diss} \sim L/V$, which is less than $\sim 10$ s 
for a length of $L \sim 10^8$ cm and $V = 1.5\times10^7$ cm s$^{-1}$. 
This implies that the heating rate would be around 10~eV per particle per second, but it declines quickly with height. 
It could approach 1000~eV per particle when integrated over 384 seconds, which would be compatible with our measurements. 
However, some energy source would have to supply energy to maintain the turbulence, maybe by merging plasmoids. 

\citet{lee17} investigated the heating rates for an erupting flux rope for the 2012 January 27 event. They found the heating rates $\sim$ 0.1 erg cm$^{-3}$ s$^{-1}$ at the beginning of the eruption. For a density of $\sim$ 10$^{10}$ cm$^{-3}$, this corresponds to $\sim$ 6~eV per particle per second, which is comparable to the heating rates in our model ($\sim 10$ eV per particle per second). This consistency supports our interpretation of the continuous heating term. Previous studies have emphasized the need for continuous heating during CME eruption to maintain high temperatures and match observations from the Ultraviolet Coronagraph Spectrometer onboard {\it{Solar and Heliospheric Observatory}} \citep{akmal01, ciaravella01, lee09, murphy11}.  

\subsection{The nature of the blob and possible heating mechanisms}

We have referred to the feature that we tracked as a 'blob' to avoid a
biased interpretation. It is tempting to identify the blob as a plasmoid,
because rapid reconnection requires that the current sheet break up into small,
discrete X-points and plasmoids \citep[e.g.,][]{loureiro2007, uzdensky2010, 
bhattacharjee2009, shibata2014}.  
While those plasmoids must be very small to enable rapid reconnection, 
they can merge to form much larger structures as they move along the current sheet  
\citep[e.g.,][]{barta2008, fermo2010, shen13}. 

On the other hand, UVCS observations
of post-flare plasma sheets showed cool blobs apparently moving along
the sides of the current sheet \citep{bemporad2003, ciaravella2008}. 
Because those blobs were so much cooler than the plasma
sheet, appearing in the Lyman lines and O VI rather than [Fe XVIII], they were
attributed to CME-like ejections resulting from secondary reconnection events 
triggered by the main flare.  
Blobs can also be seen in white light observations of current sheet.
While white light observations do not reveal temperatures
to help discriminate between features inside and outside
of the current sheet, a series of repeated outflow features, as in 
\citet{savage10, schanche2016, kim2020, lee20, patel20} seems likely to be
associated with plasmoids.

The temperature of the blob we analyze here 
closely matches the temperature of the larger plasma sheet as determined
by \citet{warren18, imada21, gou24, Kittrell25}, so it is far more likely to be feature of
the main reconnection flow. However, that does not guarantee that it is an
example of the classic plasmoid expected from theory. The theories begin 
with uniform field structures, and a complex field and density structure in the 
pre-event corona could perhaps produce the observed blob. 

The heating mechanism is even more difficult to determine. While we have
measured the amount of heat deposited as the blob moves up the current sheet, we
have little other information. Thermal conduction is unlikely to play a role, since
the temperatures of the blob and the plasma sheet match so closely. Non-thermal
particles might deposit energy in the blob, but that possibility is difficult to quantify. 
It is also possible that waves and turbulence generated in the main X-line
overtake the blob and deposit energy. In the plasmoid model of reconnection, 
reconnection at  the X-points on either side of the plasmoid converts most of 
the magnetic energy into kinetic energy, which could be deposited in the 
plasmoid by shocks or mixing. However, much of the plasmoid heating seen in 
MHD models results from reconnection during plasmoid mergers \citep{ye19}, 
and that is a plausible mechanism for the heating.

\section{Conclusion}

We investigated NEI models incorporating nonthermal electron velocity distributions to interpret EUV observations of the plasma sheet formed during the particularly well observed flux rope eruption on 2017 September 10. Using a covariance-based $\chi^2$ analysis and a $\Delta \chi^2$ confidence criterion, we explored the parameter space defined by the impulsively heated temperature, the continuous heating rate, and $\kappa$.

The statistically acceptable models require impulsive heating temperatures above approximately 20–25~MK across all decay scenarios.
High temperatures ($HT \sim 50$~MK) are allowed only at very low $\kappa$ values ($\kappa = 2$), indicating that the presence of a sufficiently strong suprathermal tail is required by the models for such temperatures to be consistent with the observations. Impulsive heating dominates the energy budget in all statistically acceptable models. The continuous heating contributes about 6\%–50\% of the initial impulsive energy input. 

Overall, our results highlight the importance of suprathermal electron populations and NEI effects in interpreting EUV observations of the plasma sheet. 
The inferred $\kappa$ values should be regarded as effective, time-averaged parameters, and further progress will require a fully self-consistent treatment coupling time-dependent electron distributions with hydrodynamic evolution. 

%% Also note that the akcnowlodgment environment does not support long amounts of text. If you have a lot of people and institutions to acknowledge, do not use this command. Instead, create a new \section{Acknowledgments}.
\begin{acknowledgments}

Dr. Stephen Kahler had passed away during this work. We honor his dedication, passion for this work, and lifetime achievements in solar and space physics. 
CHIANTI is a collaborative project involving George Mason University, the University of Michigan (USA), University of Cambridge (UK) and NASA Goddard Space Flight Center (USA). We used ChatGPT (OpenAI) to assist with English editing and with improving the clarity of the description of the statistical methodology. All analyses, calculations, and scientific interpretations were performed and verified by the authors.
This work was supported by the Air Force Office of Scientific Research under award number FA2386-20-1-4031, the National Research Foundation of Korea(NRF) grant funded by the Korea government(MSIT) (No. NRF-2020R1I1A1A01071814, RS-2023-NR076516, RS-2026-25488141), the Korea Astronomy and Space Science Institute (KASI) grants 2026-1-830-05 and 2026-1-853-01, and by NASA Grant 80NSSC19K0853 to the Smithsonian Astrophysical Observatory. K. Reeves partially supported by NSF grant AGS-2334929.
\end{acknowledgments}

\bibliography{currentsheet1}{}
\bibliographystyle{aasjournal}

\begin{table}[t]
\centering
\caption{Parameters adopted in the covariance-based fitting.}
\label{tab:dpar}
\begin{tabular}{lll}
\hline
Parameter & Value & Description \\
\hline
$g_0$ & 0.03 & Fractional amplitude of the signal-proportional band-correlated term ($\mathbf{C}_{\rm band}$) \\
$\rho_{\rm band}$ & 0.6 & Inter-band correlation coefficient for $\mathbf{C}_{\rm band}$ (same-time only) \\
$\alpha_{\rm sys}$ & 0.4 & Scaling factor for the correlated systematic component ($\mathbf{C}_{\rm sys}$) \\
$\rho_{\rm sys}$ & 0.6 & Inter-band correlation coefficient for $\mathbf{C}_{\rm sys}$ (same-time only) \\
$\sigma_{ab}$ & 0.5 & Constraint width on the band normalization factors $a_{\mathrm{band}}$ \\
\hline
\end{tabular}
\end{table}

\begin{deluxetable}{ccccc}
\tablecaption{Statistically acceptable models satisfying $\Delta\chi^2 < 8.02$
(95.4\% confidence region for three parameters)\label{tab:rmodels}}
\tablehead{
\colhead{\shortstack{Decay\\(\%)}} &
\colhead{\shortstack{HT\\(MK)}} &
\colhead{$\kappa$} &
\colhead{\shortstack{HR\\(erg cm$^{-3}$ s$^{-1}$)}} &
\colhead{\shortstack{Continuous\\ / Impulsive}}
}
\startdata
10 & 25 & 15--100 & 0.4 & 0.39 \\
10 & 32 & 4--6    & 0.2--0.4 & 0.16--0.31 \\
10 & 40 & 3       & 0.12--0.20 & 0.07--0.12 \\
\hline
66 & 25 & 5--6    & 0.8 & 0.50 \\
66 & 32 & 3--6    & 0.4--0.8 & 0.2--0.4 \\
66 & 40 & 3       & 0.16--0.4 & 0.06--0.16 \\
66 & 50 & 2       & 1.2 & 0.38 \\
\hline
80 & 25 & 5--12   & 0.8 & 0.43 \\
80 & 32 & 3--6    & 0.4--0.8 & 0.2--0.34 \\
80 & 40 & 3       & 0.2--0.4 & 0.07--0.13 \\
80 & 50 & 2       & 1.2 & 0.32 \\
\enddata
\end{deluxetable}

\begin{deluxetable}{ccccccccc}
\label{tab:bmodels}
\tablecaption{Representative models for three decay senarios}
\tablehead{
\colhead{Decay (\%)} &
\colhead{$\kappa$} &
\colhead{$HR$ (erg cm$^{-3}$ s$^{-1}$)} &
\colhead{$HT$ (MK)} &
\colhead{$\chi_r^2$} &
\colhead{$a_{94}$} &
\colhead{$a_{131}$} &
\colhead{$a_{193}$} &
\colhead{$a_{211}$}
}
\startdata
10 & 4 & 0.4 & 32 & 0.96 & 3.1 & 0.7 & 0.9 & 1.4 \\
10 & 100 & 0.4 & 25 & 1.08 &  3.0 & 0.7 & 0.9 & 1.3 \\
10 & 3 & 0.12 & 40 &  1.10 &  3.1 & 0.8 & 1.0 & 1.4 \\
\hline
66 & 3 & 0.8 & 32 & 0.99 & 2.6 & 0.6 & 1.0 & 1.4 \\
66 & 5 & 0.8 & 25 & 1.09 & 2.4 & 0.5 & 0.9 & 1.3 \\
66 & 2 & 1.2 & 50 & 1.14 & 2.6 & 0.6 & 1.1 & 1.4 \\
\hline
80 & 3 & 0.8 & 32 & 0.95 & 2.5 & 0.5 & 1.0 & 1.4 \\
80 & 2 & 1.2 & 50 & 1.08 & 2.5 & 0.6 & 1.1 & 1.4 \\
80 & 12 & 0.8 & 25 & 1.08 & 3.1 & 0.7 & 0.9 & 1.3 
\enddata
\end{deluxetable}

%Fig 1

%Fig 2

%Fig 3

\end{document}